\documentclass[11pt]{article}

\usepackage{graphicx,epsfig}
\usepackage{multicol}
\usepackage{amsmath,amssymb,cite,color,hyperref}
\newcommand{\be}{\begin{equation}}
\newcommand{\ee}{\end{equation}}
\newcommand{\bea}{\begin{eqnarray}}
\newcommand{\eea}{\end{eqnarray}}
\newcommand{\bi}{\begin{itemize}}
\newcommand{\ei}{\end{itemize}}
\newcommand{\ben}{\begin{enumerate}}
\newcommand{\een}{\end{enumerate}}

\newcommand{\lp}{\left(}
\newcommand{\rp}{\right)}

\def\frac#1#2{{{#1}\over {#2}}}
\def\gsim{\mathrel{\rlap{\lower4pt\hbox{\hskip1pt$\sim$}}
    \raise1pt\hbox{$>$}}}         
\def\lsim{\mathrel{\rlap{\lower4pt\hbox{\hskip1pt$\sim$}}
    \raise1pt\hbox{$<$}}}         

\newcommand{\draft}[1]{}

\definecolor{grey}{rgb}{0.5,0.5,0.5}

\usepackage{cite}
\usepackage{hyperref}
\usepackage{url}

\begin{document}
\begin{figure}[h!]
\epsfig{width=0.32\textwidth,figure=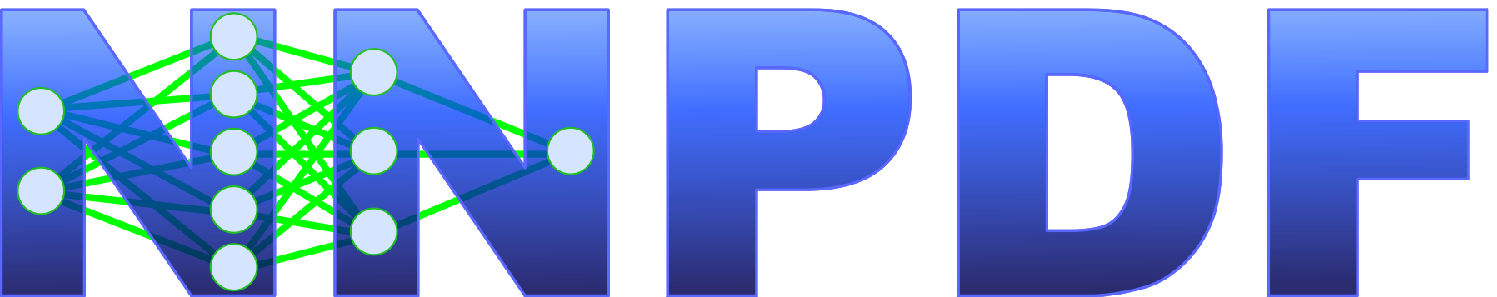}
\end{figure}

\begin{flushright}
CERN-PH-TH/2013-035\\
Edinburgh 2013/02\\
FR-PHENO-2013-003\\
IFUM-1008-FT\\
TTK-13-06\\
\end{flushright}

\begin{center}
{\large\bf Theoretical issues in PDF determination and associated uncertainties}
\vspace{0.6cm}

{\bf  The NNPDF Collaboration:}\\

Richard~D.~Ball,$^{1}$ Valerio Bertone,$^2$
Luigi~Del~Debbio,$^1$ Stefano~Forte,$^3$ \\ 
Alberto~Guffanti,$^{4}$ Juan~Rojo$^5$  and Maria~Ubiali.$^6$

\vspace{.3cm}
{\it ~$^1$ Tait Institute, University of Edinburgh,\\
JCMB, KB, Mayfield Rd, Edinburgh EH9 3JZ, Scotland \\
~$^2$  Physikalisches Institut, Albert-Ludwigs-Universit\"at Freiburg,\\ 
Hermann-Herder-Stra\ss e 3, D-79104 Freiburg i. B., Germany  \\
~$^3$ Dipartimento di Fisica, Universit\`a di Milano and
INFN, Sezione di Milano,\\ Via Celoria 16, I-20133 Milano, Italy\\
~$^4$ The Niels Bohr International Academy and Discovery Center, \\
The Niels Bohr Institute, Blegdamsvej 17, DK-2100 Copenhagen, Denmark\\
~$^5$ PH Department, TH Unit, CERN, CH-1211 Geneva 23, Switzerland \\
~$^6$ Institut f\"ur Theoretische Teilchenphysik und Kosmologie, RWTH 
Aachen University,\\ D-52056 Aachen, Germany\\}
\end{center}   

\vspace{0.2cm}

\begin{center}
{\bf \large Abstract}
\end{center}
We study several sources of theoretical uncertainty in the
determination of parton distributions (PDFs) which may affect current PDF
sets used for precision physics at the Large
Hadron Collider, and explain discrepancies
between them.
We consider in particular the use of fixed-flavor versus
variable-flavor number renormalization schemes, higher twist
corrections, and nuclear corrections. We perform our study in the
framework of the NNPDF2.3 global PDF determination, by 
quantifying in each case the impact of different theoretical assumptions
on the output PDFs.  We also study in each case the implications
for benchmark cross sections at the LHC.
We find  that the impact in a global fit of 
a  fixed-flavor number scheme is substantial, the
impact of higher twists is negligible, and the impact of nuclear
corrections is moderate and circumscribed.

\clearpage

Precision physics at the LHC requires ever better  
estimates of parton distribution (PDF) uncertainties (see e.g.~\cite{Forte:2013wc}). At present PDF
uncertainties do not include all sources of theoretical uncertainty: they
only reflect the uncertainty of the underlying data, and (possibly) in the procedure used in the
PDF determination, but not the effect of the some of the necessary theoretical approximations. 
However, these published PDF uncertainties are now rapidly decreasing because of 
the availability of abundant and precise new data from the LHC. Hence, theoretical uncertainties 
are soon going to become significant and in certain cases even dominant. Indeed, this might
already be sometimes the case: a recent benchmarking of the dependence on PDFs of 
predictions  for several LHC processes~\cite{Ball:2012wy} shows that in some  cases predictions
obtained using different PDF sets disagree by a sizable amount on the scale of PDF uncertainties. 
It is then natural to ask whether some known differences in the theoretical approach used in 
different PDF extractions may explain these differences.

There are two sources of  theoretical uncertainty on which there is currently some, albeit partial, 
knowledge. The first is the dependence on the perturbative order. Since all PDF
sets~\cite{Forte:2013wc} are now available at NLO and NNLO (and most also at LO), the uncertainty 
on the NLO results can be determined exactly, and that on the NNLO result can be at least in principle
estimated from the behavior of the perturbative series~\cite{Cacciari:2011ze}. 
The second is the dependence on the matching scheme used to include heavy quark masses. Most PDF 
fitting groups use a so-called general-mass variable-flavor number (GM-VFN) scheme to combine fixed 
order  contributions computed with full inclusion of heavy quark masses with all-order resummation of 
contributions due to perturbative evolution in which heavy quarks are treated as massless partons. 
Several ways of doing so used by various PDF fitting groups, which differ by subleading terms, have 
been compared and benchmarked in Ref.~\cite{LHhq}. However, some PDF fitting groups  use a 
fixed-flavor number (FFN) scheme, where only the three lightest 
flavors and antiflavours are treated as massless partons and enter QCD evolution equations, while the
contributions of heavy quarks are included in partonic cross sections. 
There are indications~\cite{Thorne:2012az} that this choice may explain some, or perhaps even most, 
of the differences between PDF sets.  
Hence this issue deserves further investigation. 

There are two further obvious, potentially large, sources of theoretical uncertainty in PDF fits. 
The first is related 
to the treatment of the medium energy region, where power-suppressed (higher-twist) contributions   
to the Wilson expansion may be relevant, especially for deep-inelastic scattering (DIS) data, 
the kinematic coverage of  which sometimes extends to relatively low scales, not much above the 
nucleon mass. 
While the lowest-scale, potentially dangerous DIS data are usually excluded from PDF determinations 
by suitably 
chosen kinematic cuts, the potential impact of such data (and thus in particular the dependence on 
the choice of 
kinematic cuts) needs to be studied systematically. The second source of uncertainty is related to 
the fact that a sizable fraction of the DIS data (and also some 
Drell-Yan data) are obtained using nuclear targets: deuterium for charged-lepton DIS, and heavy nuclei for 
neutrino DIS. These data are crucial for the separation of light flavors, and it has been suggested
recently~\cite{Accardi:2011fa,Owens:2012bv} that corrections due to nuclear structure may have a significant 
impact on the extraction of PDFs.

In this paper we will consider these three, possibly dominant, sources of theoretical uncertainties 
on PDFs: the use of a FFN scheme, the impact of higher twist terms, and the impact of nuclear corrections. In each case, we 
will repeat the NNPDF2.3 NNLO PDF  fit~\cite{Ball:2012cx} by varying the way these effects are treated, 
and will compare the results both in terms of their impact on PDFs and the quality of the fit, 
and also by checking 
their impact on standard LHC observables.
  
\begin{figure}
    \begin{center}
\includegraphics[width=0.90\textwidth]{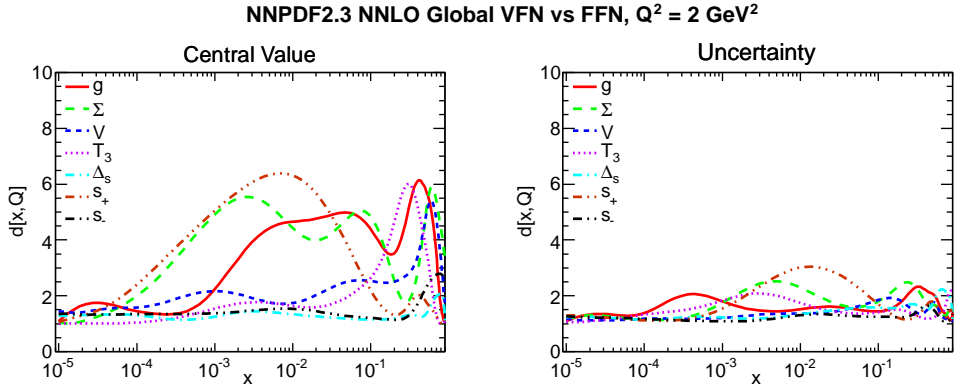}
      \end{center}
     \caption{\small 
    \label{fig:distances-ffn-lowQ2} 
Distances between the default NNPDF2.3 PDFs, and PDFs obtained treating DIS data in a FFN scheme. 
Distances are shown at the scale  $Q^2=2$ GeV$^2$ at which PDFs are parametrized, both between 
central values (left) and uncertainties (right).}
\end{figure}

\bigskip\noindent
{\bf FFN Schemes}

We first discuss the impact of the use of the FFN scheme to treat heavy flavor contributions. In the default NNPDF2.3 
fit heavy quark mass effects in deep-inelastic structure functions are included using the FONLL-C scheme (see
Refs.~\cite{Ball:2012cx,Forte:2010ta}). In this study we have
performed a new fit in which all  data for DIS 
structure functions are treated using a FFN scheme, 
while other (hadronic) data are treated in the VFN scheme, with 
massless heavy quarks.
In the FFN fit, we take, for the DIS data, $\alpha_s^{(n_f=3)}(m_c)=0.3680$ at $m_c^2=2$ GeV$^2$, which corresponds 
to $\alpha_s^{(n_f=5)}(M_Z)=0.119$, and  $\alpha_s^{(n_f=3)}(M_Z)=0.1061$. For the hadronic data, we also use 
$\alpha_s^{(n_f=5)}(M_Z)=0.119$. The fits are performed at NNLO, using
$O(\alpha_s^2)$ massive coefficient functions for charm (namely, the
same massive charm terms as in the FONLL-C scheme). 

The rationale for only treating DIS data in an FFN scheme in our study 
is that the  use of a FFN scheme has been
advocated~\cite{Alekhin:2009ni,Alekhin:2012ig} mostly in conjunction
to the inclusion of heavy quark mass terms in deep-inelastic heavy
quark production. 
Heavy quark mass corrections to inclusive hadronic
processes used in PDF determination are usually not included (though
this could be done also in a VFN scheme~\cite{Cacciari:1998it} using
the  FONLL method used by NNPDF),  so nothing is to be learnt  by using a FFN 
scheme in the description of these data. 

In Fig.~\ref{fig:distances-ffn-lowQ2} we show the distances between central 
values and uncertainties of PDFs thus determined, and the default NNPDF2.3 PDFs, at the initial scale 
$Q_0^2=2$~GeV$^2$. 
The distance
$d(x,Q^2)$ between a pair of replica samples for a certain PDF  
at a given value of $x$ and
$Q^2$
(as defined in Appendix A of
Ref.~\cite{Ball:2010de})   
is basically the difference of the means of the two samples, 
in units of the standard deviation of
the mean (distance between central values), or the difference 
of their standard deviations in units of the standard
deviation of the standard deviation  (distance between uncertainties). The definition entails that if we
compare two different samples of $N_{\rm rep}$ replicas, each extracted
from the same distribution, then on average $d=1$, while if the two
samples are extracted from two distributions whose means differ by one
standard deviation, then on average $d=\sqrt{N_{\rm rep}}$, the
difference being due to the fact that the standard deviation of the
mean scales as $1/\sqrt{N_{\rm rep}}$. 
So  $d\sim1$ corresponds to statistically equivalent 
PDFs, while $d\sim 10$ (with $N_{\rm rep}=100$ replicas) corresponds to statistically inequivalent PDFs which differ 
by one sigma. 

It is clear from Fig.~\ref{fig:distances-ffn-lowQ2}  that the two fits are inequivalent, but
differences are moderate, at the half sigma level or so, 
with the change being observed in central values, but with no significant change in uncertainties. 
The PDF which varies most is the gluon: this is easily understandable given that the gluon is determined mostly by scaling 
violations, which are different in the FFN case.

\begin{figure}
    \begin{center}
\includegraphics[width=0.90\textwidth]{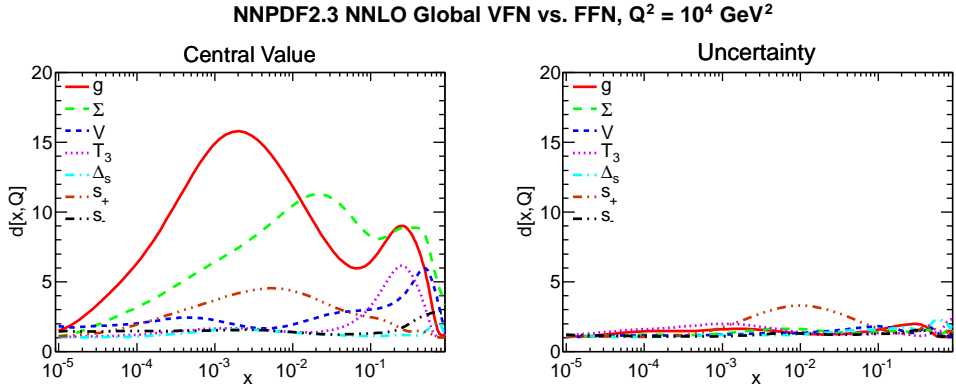}
      \end{center}
     \caption{\small 
    \label{fig:distances-ffn-highQ2}
Same as Fig.~\ref{fig:distances-ffn-lowQ2}, but at $Q^2=10^4$
GeV$^2$. All PDFs have been evolved upwards using the same standard
(VFN) evolution equations.}
\end{figure}

\begin{figure}[t]
    \begin{center}
\includegraphics[width=0.43\textwidth]{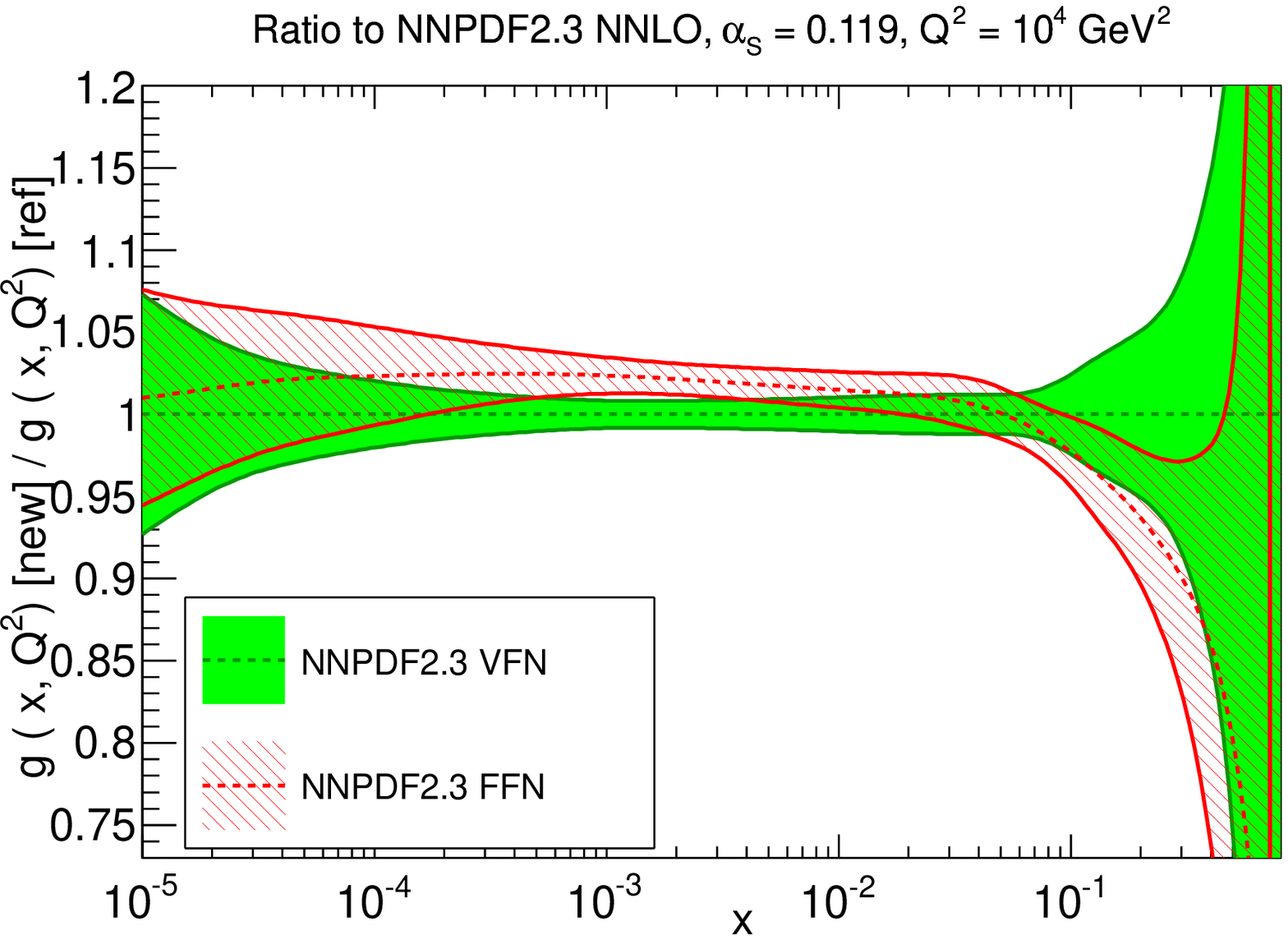}
\includegraphics[width=0.43\textwidth]{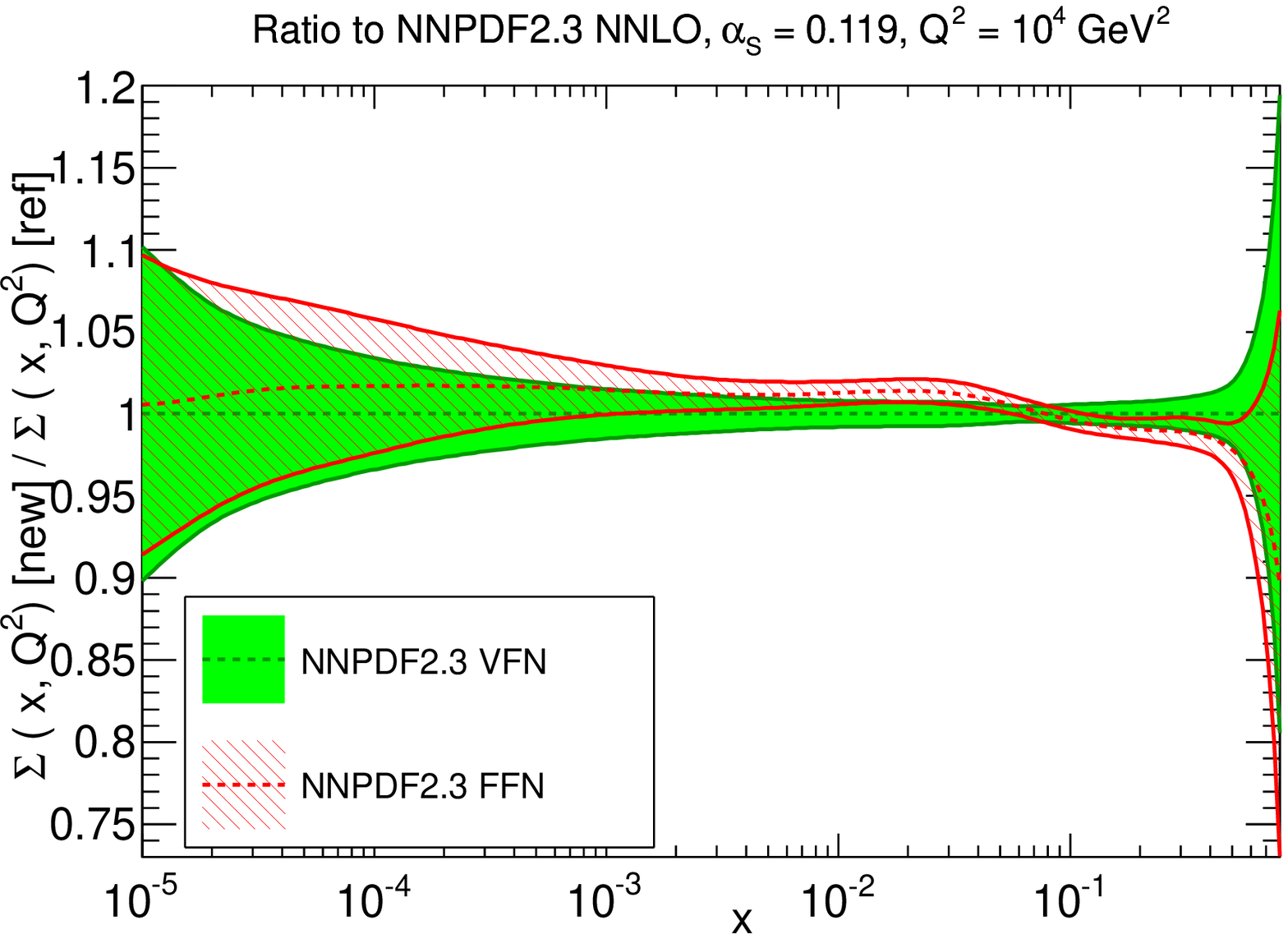}\\
\includegraphics[width=0.43\textwidth]{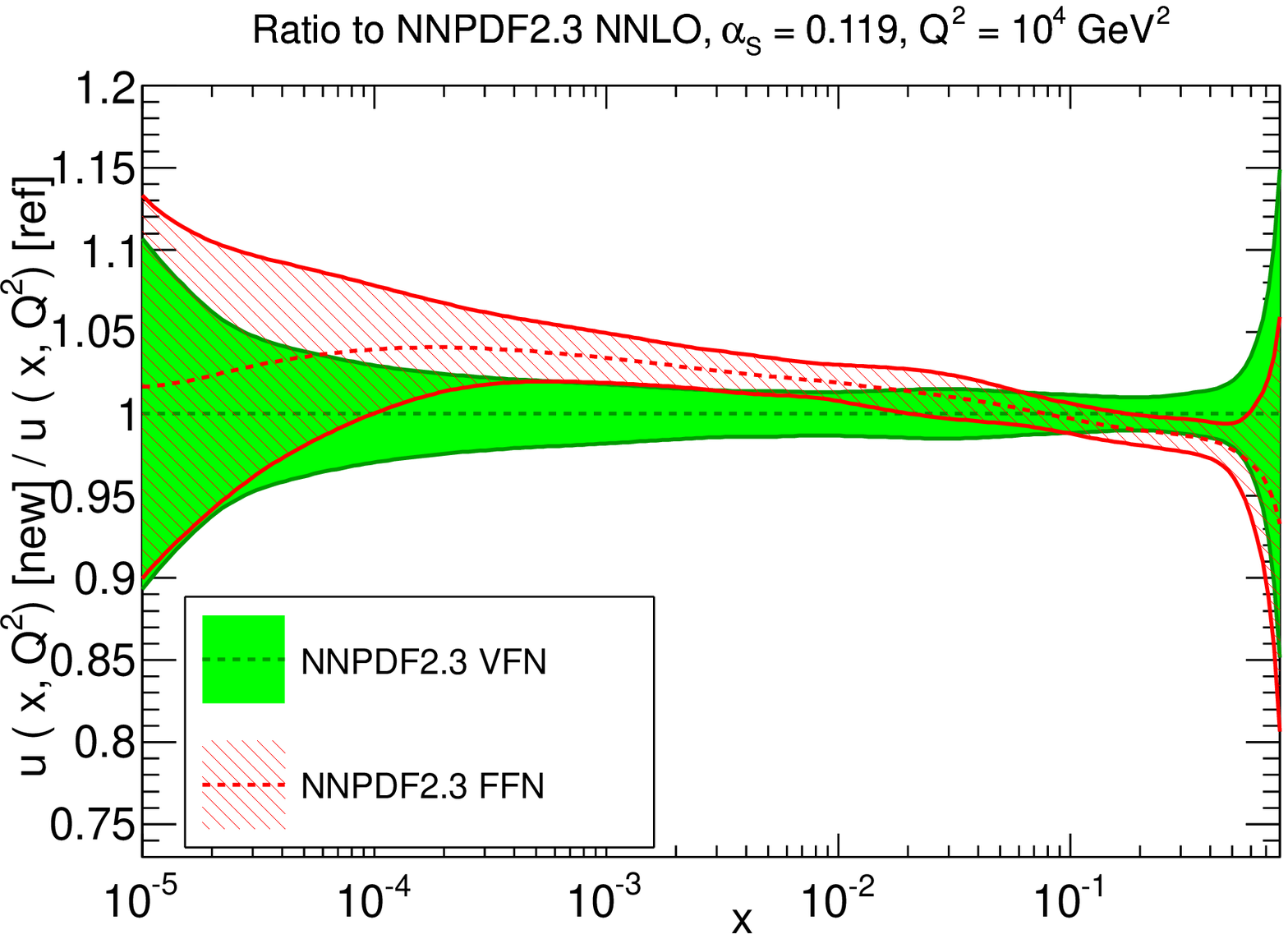}
\includegraphics[width=0.43\textwidth]{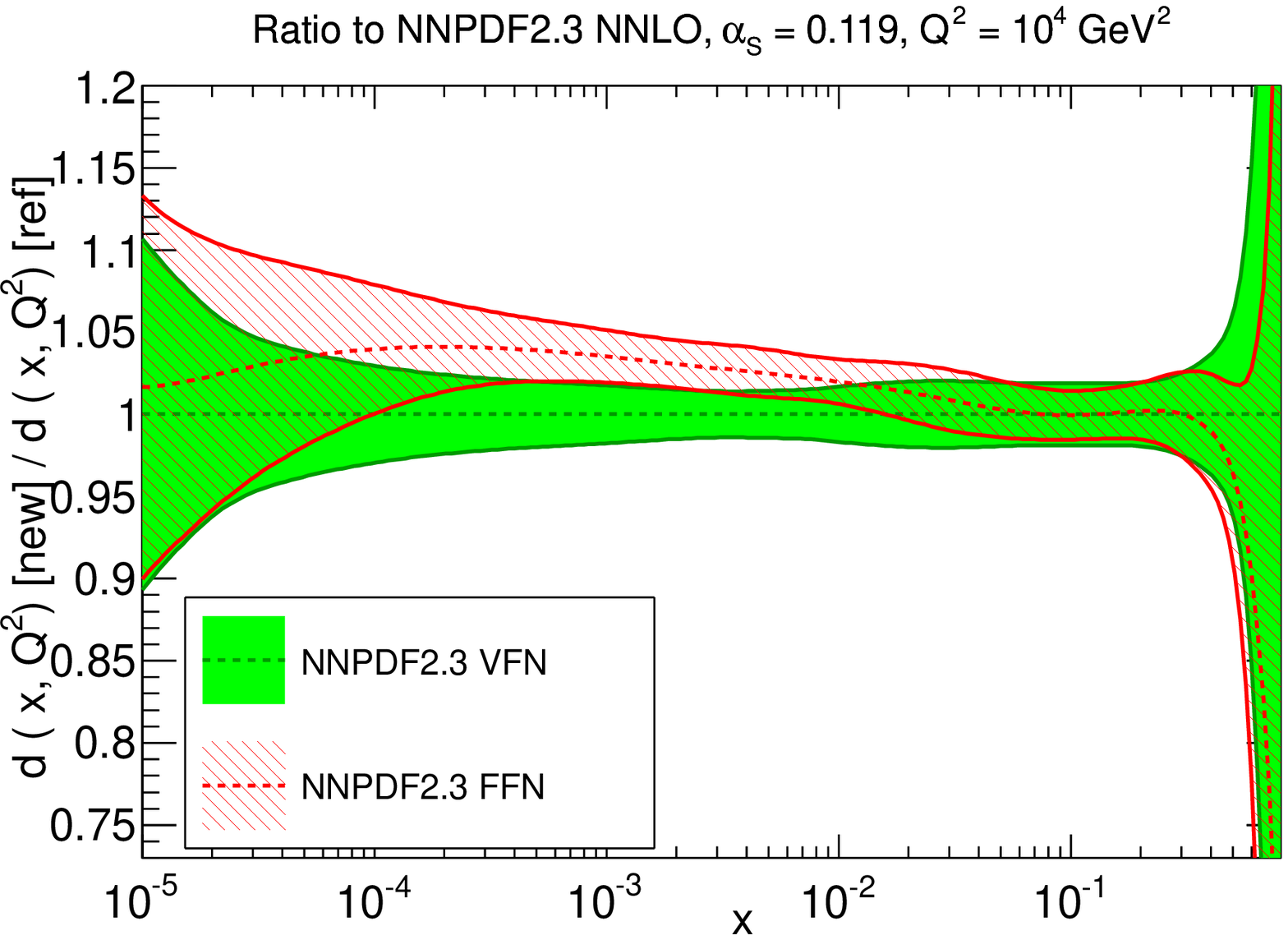}\\
      \end{center}
     \caption{\small 
    \label{fig:pdfplots_ffn}
Comparison between FFN and default  NNPDF2.3 PDFs,
displayed as a ratio to the default
NNPDF2.3 at
 $Q^2=10^4$~GeV$^2$: gluon (top left), quark singlet (top right), up
(bottom left) and down (bottom right).}
\end{figure}

For collider physics applications, the initial PDFs, whether determined in a GM-VFN scheme, or  
in a FFN scheme~\cite{Alekhin:2009ni,Alekhin:2012ig}, are evolved upwards using the usual VFN 
evolution equations. 
It turns out that this evolution amplifies differences between GM-VFN and FFN PDFs. 
The amplification is demonstrated in Fig.~\ref{fig:distances-ffn-highQ2}, where the same distances of
Fig.~\ref{fig:distances-ffn-highQ2} are shown, but now at the scale $Q^2=10^{4}$~GeV$^2$ (relevant e.g. for $W$, $Z$ 
or Higgs production). While uncertainties are still unchanged, central values for some PDFs (specifically the gluon and 
the quark singlet) now differ by more than one sigma. 
The fact that differences become larger when evolving to higher scales can be understood as a consequence of the fact 
that differences in the large $x$ region (where uncertainties are large) at low scale lead upon evolution to high scale 
to differences in the small $x$ region, where uncertainties are relatively small.
In Fig.~\ref{fig:pdfplots_ffn} the PDFs that change most with respect to the standard NNPDF2.3 ones when adopting a 
FFN scheme are compared to their default counterparts, shown as a ratio to NNPDF2.3.

Having ascertained that the impact of choosing a FFN scheme is not negligible, we next ask 
whether a theoretical uncertainty of order of the difference between PDFs extracted in the 
FFN and GM-VFN schemes is being neglected when results are presented in a particular scheme. 
In order to understand if this is the case, we 
have studied the fit quality in the two fits. In Tab.~\ref{tab:chi2tab} we show the  difference between the $\chi^2$ of the 
DIS data, computed using the  FFN or VFN PDF sets both for the total DIS dataset, or for the subset of the combined HERA-I 
data~\cite{H1:2009wt}, in various kinematic regions. The so-called  `experimental' definition of the $\chi^2$ (to be found in  
the Appendix of Ref~\cite{Ball:2012wy}) is used. 
The contribution to the difference from HERA DIS data is shown separately, and in each case the number of data points 
is given. Note that the default NNPDF2.3 cut on the final state invariant mass $W^2>12.5$~GeV$^2$ is always imposed. 
The difference is always positive, indicating that the fit quality is
always worse in the FFN case than it is with the default GM-VFN scheme.
The difference in $\chi^2$ is also positive for all the remaining data in the global fit, and is equal to about 20 units for about 
750 data points. Hence, the FFN PDFs provide a worse fit to the global dataset, and especially a worse fit to DIS data.

\begin{table}[t]
\centering
\small
 \begin{tabular}{c|c|c|c|c|c|c|c}
 \hline
 $x_{\rm min}$ & $x_{\rm max}$ &$Q^2_{\rm min}$ (GeV) & $Q^2_{\rm max}$
(GeV) &$\Delta \chi^2$ (DIS)& $N_{\rm dat}^{\rm DIS}$ & $\Delta \chi^2$ (HERA-I)& $N_{\rm dat}^{\rm hera-I}$ \\ 
 \hline
 \hline
$  4 \cdot 10^{-5}  $&$   1 $ &$   3  $&$    10^6  $&      72.2  &  2936  &      77.1  &   592  \\ 
$  4 \cdot 10^{-5}  $&$  0.1 $ &$   3  $&$    10^6  $&      87.1  &  1055  &      67.8  &   405  \\ 
$ 4 \cdot  10^{-5}  $&$    0.01 $ &$  3  $&$    10^6  $&      40.9  &   422  &      17.8  &   202  \\ 
$ 4 \cdot  10^{-5}  $&$   1 $ &$   10  $&$    10^6  $&      53.6  &  2109  &      76.4  &   537  \\ 
$ 4 \cdot  10^{-5}  $&$   1 $ &$   100  $&$    10^6  $&      91.4  &   620  &      97.7  &   412  \\ 
$ 4 \cdot  10^{-5}  $&$  0.1 $ &$   10  $&$    10^6  $&      84.9  &   583  &      67.4  &   350  \\ 
$ 4 \cdot  10^{-5}  $&$  0.1 $ &$   100  $&$    10^6  $&      87.7  &   321  &      87.1  &   227  \\ 
 \hline
 \end{tabular}
\caption{\small Difference 
$\Delta \chi^2\equiv \chi^2_{\rm FFN}-\chi^2_{\rm VFN}$ of the deep-inelastic
 data, as
  described using FFN theory with the best-fit PDFs obtained from a a
  global fit in which PDFs are treated in a FFN scheme, and using the
  default NNPDF2.3 set. Note that the numbers
shown are for the absolute $\chi^2$, not divided by the
number of data points.
The contribution from  combined HERA-I data is also shown in the last
column. Results are shown in various kinematic regions,
taking into account experimental
correlations. The number of
DIS or HERA-I
data points  after cuts is also shown in each case. The first row
corresponds to the default cuts of the NNPDF2.3 fit.
 \label{tab:chi2tab} }
\end{table}

Comparing the fit quality in different kinematic regions, one can see that the deterioration in fit quality for the FFN 
structure functions is concentrated in the region of large $Q^2\gsim100$~GeV$^2$ and small $x\lsim 0.1$ (and thus mostly the HERA data).
This may perhaps be understood in terms of the so-called `double-asymptotic scaling' properties of the 
structure function $F_2$ in this region~\cite{das}: the rise of the structure function at small $x$ has a universal logarithmic 
slope, driven by perturbative evolution,  which depends on the number of active flavors, and current HERA data are 
precise enough to see the change of slope when going above $b$ threshold (see Ref.~\cite{Caola:2010cy}, in particular 
Fig.~5). In a FFN scheme the contribution of heavy flavors to this rise is expanded out to finite order rather than being 
exponentiated to all orders, and this is likely to provide a worse description of this double scaling behaviour.

We conclude that the FFN fit is actually based on a less precise theory, in that it does not include full resummation of 
the contribution of heavy quarks to perturbative PDF evolution, and thus provides a less accurate description of the data. 
The difference between FFN and GM-VFN PDFs should, therefore, be added as a theoretical uncertainty to FFN PDFs, but 
not to GM-VFN ones, just like the difference between NLO and NNLO PDFs is part of the theoretical uncertainty on NLO 
PDFs, which disappears when going to NNLO.

\bigskip\noindent
{\bf Higher Twist}

We turn now to the study of the impact on PDF determinations of power-suppressed corrections. As is 
well known, DIS
structure function data are affected both by power corrections of kinematic origin related to the mass of the target 
(TMCs, henceforth),  as well as corrections related to higher twist contributions to the Wilson expansion. 
The former can be determined exactly in the form of an expansion in powers of $m^2_N/Q^2$, with $m_N$ the nucleon 
target mass, while the latter are of dynamical origin, and thus if included 
they must be fitted, just like the leading twist PDFs. 

Currently, TMCs are included up to $O\left(m^2_N/Q^2\right)$ in the NNPDF2.3 (and in fact in all previous NNPDF sets) 
and in the ABM11 PDF  determinations, though not in other PDF determinations such as MSTW08~\cite{Martin:2009iq},
CT10\cite{Gao:2013xoa}, and HERAPDF1.5~\cite{Radescu:2010zz,CooperSarkar:2011aa}, where they are
kept under control through suitable kinematic cuts. Dynamical higher twist corrections, on the other hand, are parametrized 
and fitted in the ABM11 PDF determination, but not in any of the other PDF sets, and in particular not in NNPDF2.3, 
where again they are kept under control by imposing a suitable kinematic cut on the invariant mass of the final state: NNPDF2.3 
removes DIS data for which $W^2\le 12.5$ GeV$^2$, and only includes
data with $Q^2\ge 3.0$~GeV$^2$. Similar cuts are adopted in other
global fits:  CT10 removes data with $W^2\le 12.25$ GeV$^2$ and only
accepts data with $Q^2\ge 4.0$~GeV$^2$; MSTW08 removes data with
$W^2\le 15.0$ GeV$^2$ and only 
accepts data with $Q^2\ge 2.0$~GeV$^2$. HERAPDF1.5 only removes data with
$Q^2<3.5$ as there are no low $W^2$ data in their dataset, while ABM11, who
fit higher twist corrections, only remove DIS data with $W^2\le 3.24$
GeV$^2$, while including all 
data with $Q^2\ge 3.24$~GeV$^2$.

\begin{figure}[h!]
    \begin{center}
\includegraphics[width=0.55\textwidth]{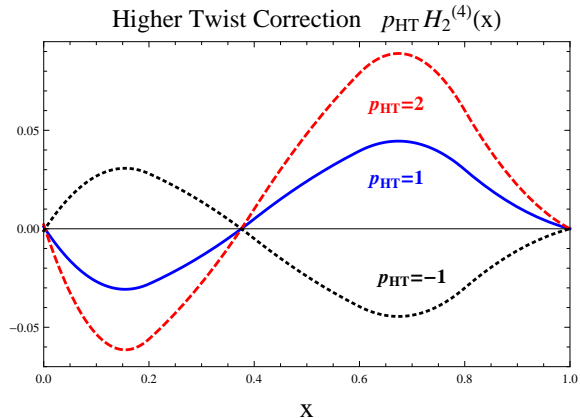}
    \end{center}
    \caption{\small 
      \label{fig:ht_corr} 
      The shape of the higher twist correction $H_2^{(4)}(x)$
    Eq.~(\ref{eq:FHT}), for the proton structure
function $F_2^p$, as determined in
    Ref.~\cite{Alekhin:2012ig}. The correction is also shown rescaled
    by a factor $p_{\rm HT}$=-1 or  $p_{\rm HT}$=2.
 }
\end{figure}

One may nevertheless worry that despite these cuts there might be a residual non-negligible uncertainty related 
to the neglect of higher twist corrections. To estimate it, we  have performed a series of fits in which the  leading 
twist computation of DIS structure functions is supplemented by a twist four correction
\be
\label{eq:FHT}
F_i^{\rm HT}(x,Q^2) = F_i^{\rm LT}(x,Q^2) + p_{\rm HT}\frac{H_i^{(4)}(x)}{Q^2} \ , 
\ee
where  $F_i^{\rm LT}(x,Q^2)$  is the leading twist NNPDF2.3 determination of the longitudinal or transverse 
structure function (including target-mass corrections), $H_i^{(4)}(x)$ is a function, with dimensions of mass squared, 
assumed to be independent of $Q^2$ (thus neglecting the logarithmic scale dependence of higher twist corrections) 
and to be taken from models or from an independent fit, and $p_{\rm HT}$ is a constant, to be used to rescale the size of 
the higher twist correction.
We have further assumed for $H_i^{(4)}(x)$ the form that was obtained in Ref.~\cite{Alekhin:2012ig} along with the ABM11 
PDF set, and we have varied the parameter $-1\le p_{\rm HT}\le 2$, i.e. we have made it twice as large, or 
reversed its sign. The shape of $p_{\rm HT}H_i^{(4)}(x)$ is displayed in Fig.~\ref{fig:ht_corr}. 

\begin{figure}[h!]
    \begin{center}
\includegraphics[width=0.90\textwidth]{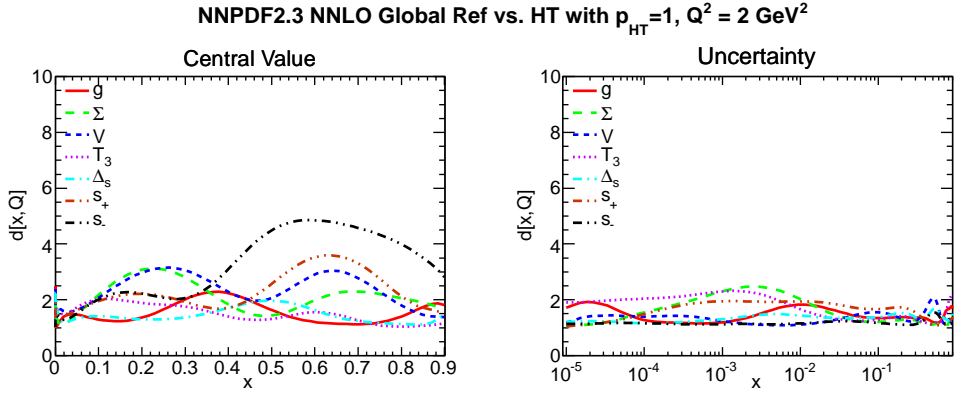}  
      \end{center}
     \caption{\small 
    \label{fig:distances-HTp1} 
Distances between PDFs determined including the higher twist correction
shown in Fig.~\ref{fig:ht_corr} with $p_{\rm HT}=1$ in
Eq.~(\ref{eq:FHT}), and the default 
 NNPDF2.3 set.}
\end{figure}

\begin{figure}[h!]
    \begin{center}
\includegraphics[width=0.45\textwidth]{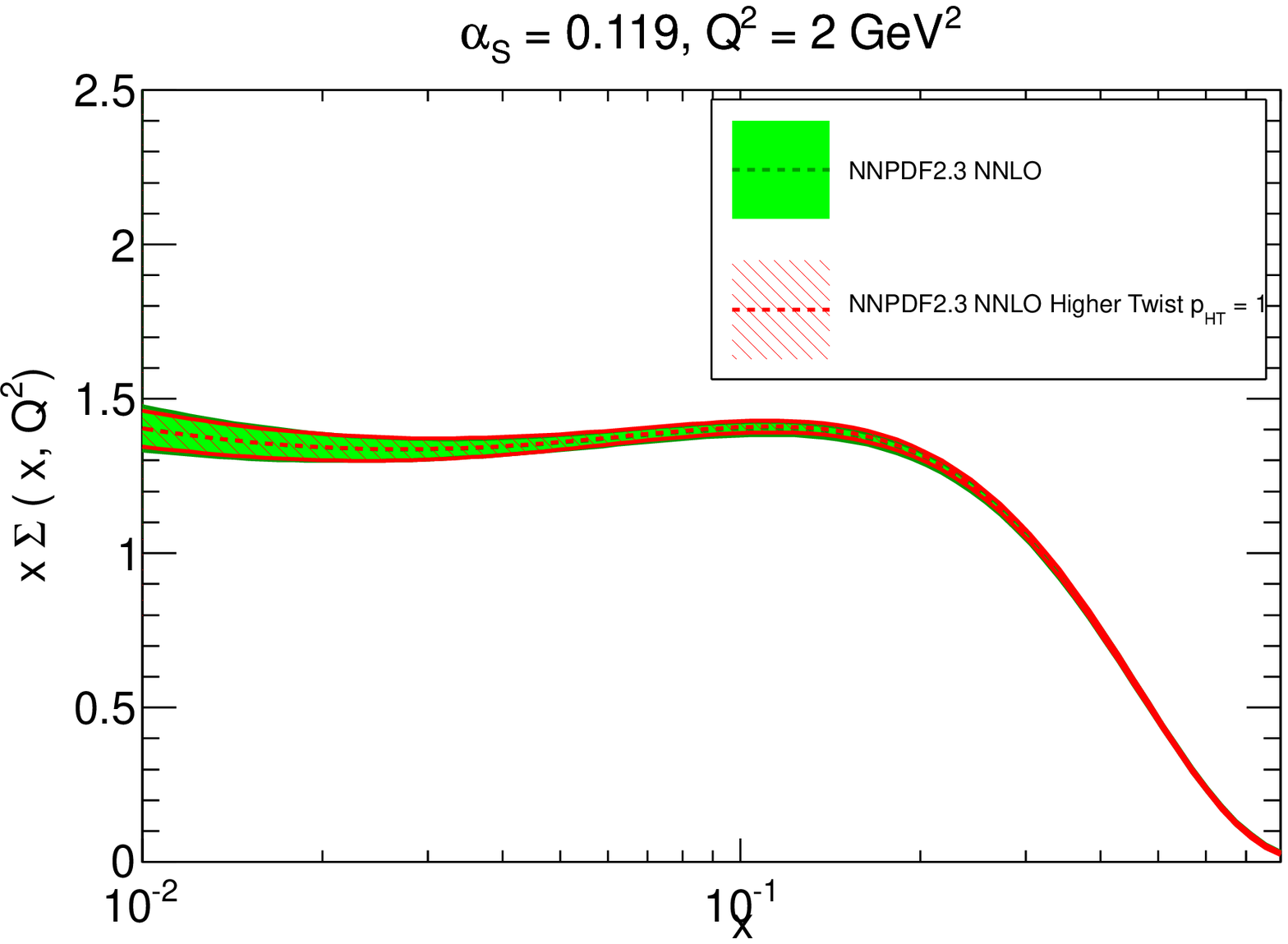}
\includegraphics[width=0.45\textwidth]{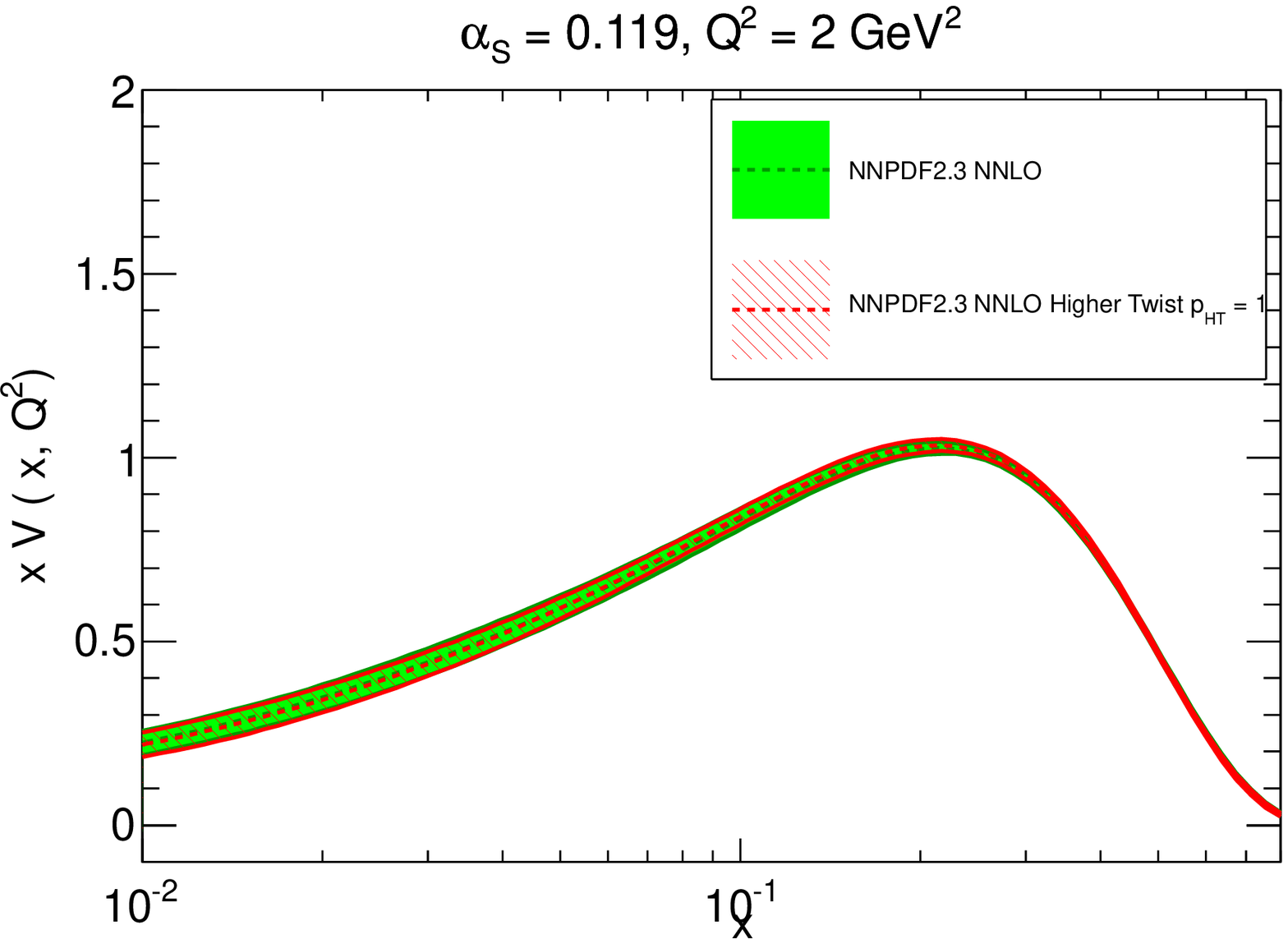}\\
\includegraphics[width=0.45\textwidth]{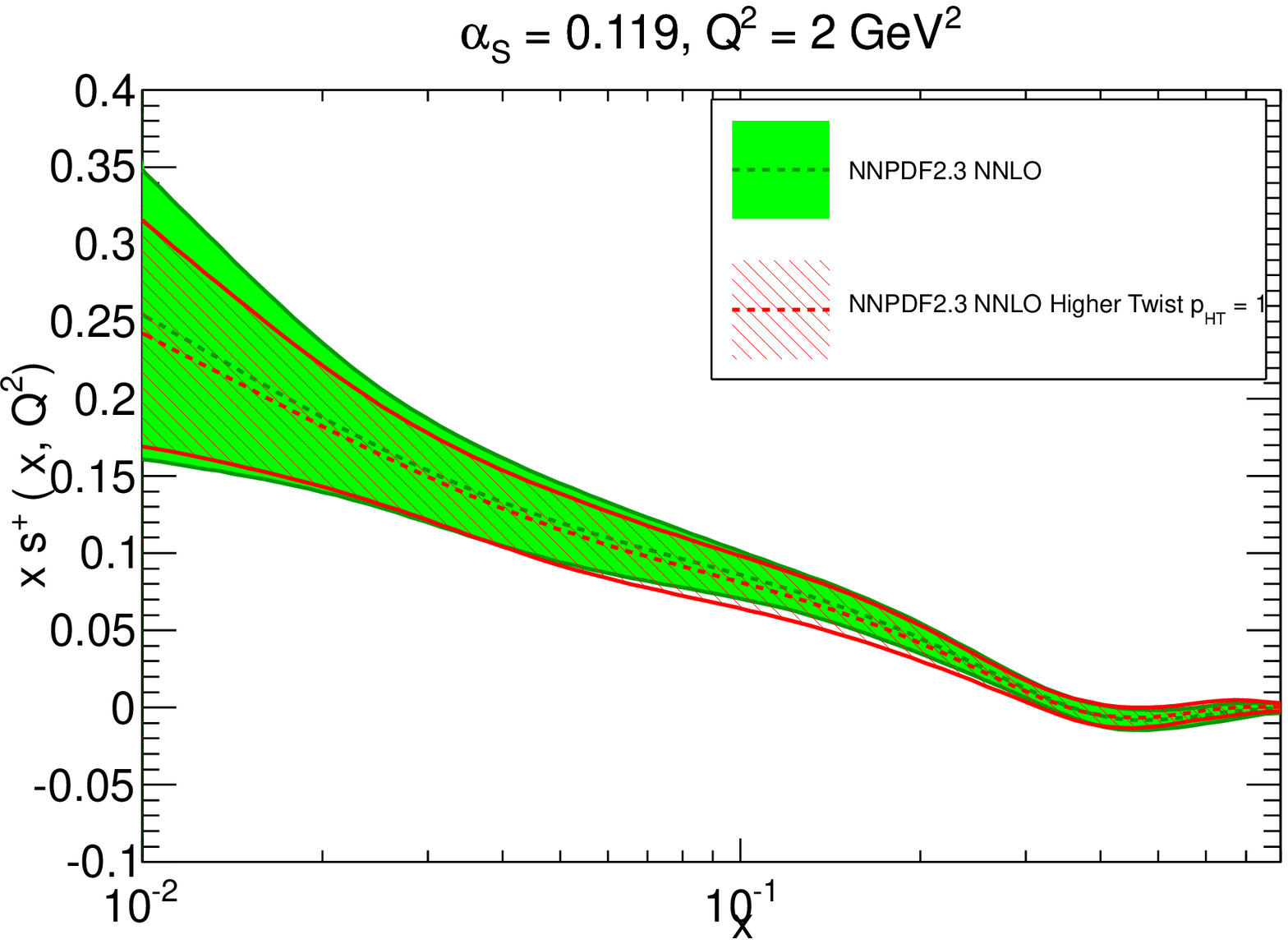}
\includegraphics[width=0.45\textwidth]{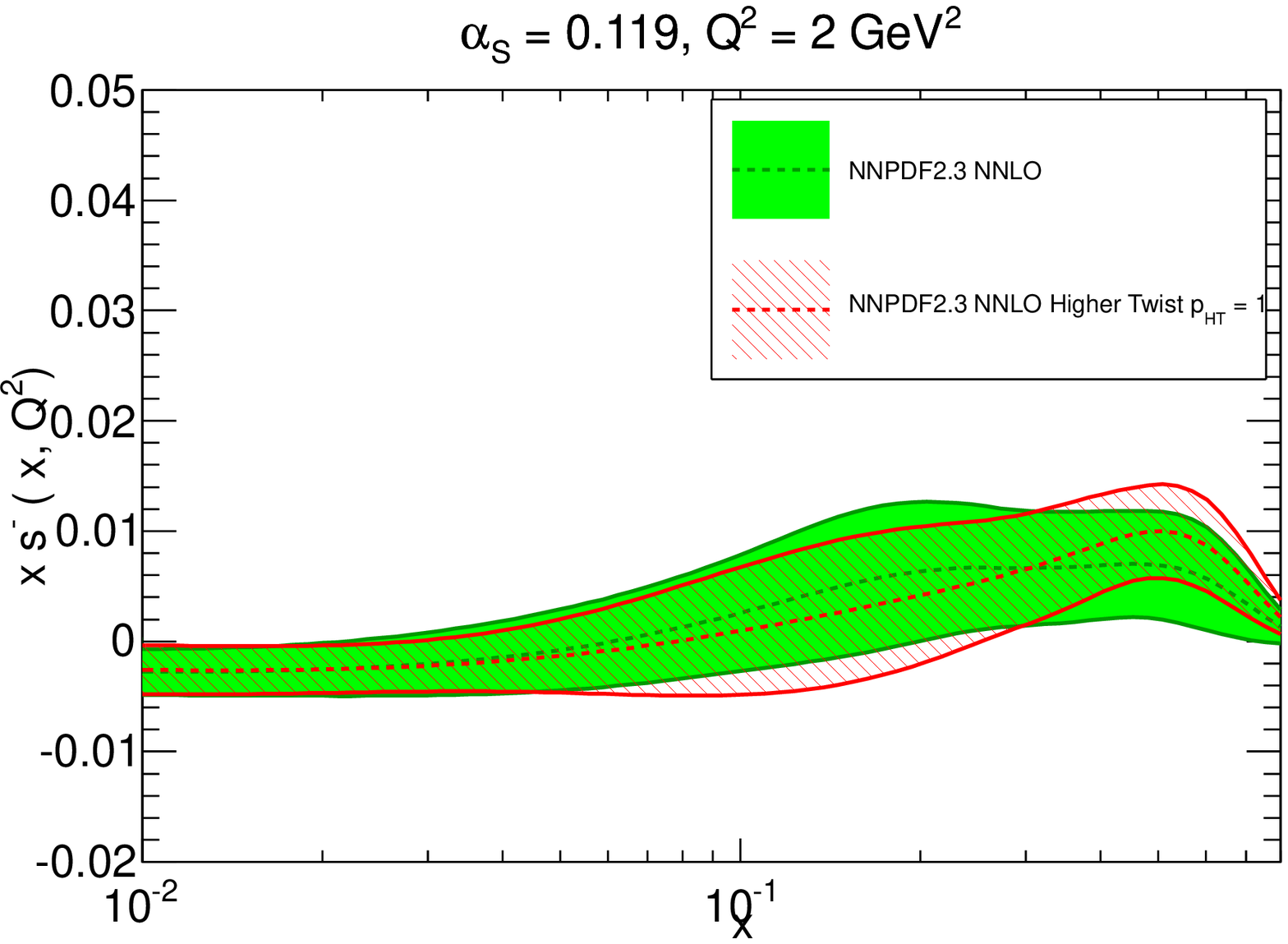}\\
      \end{center}
     \caption{\small 
    \label{fig:comp-HTp1}
Comparison of PDFs which are mostly affected by higher twist
corrections,  Eq.~(\ref{eq:FHT}) with $p_{\rm HT}=1$,
 and the default (distances were shown   in
Fig.~\ref{fig:distances-HTp1}): singlet (top left), total valence
quark (top right), strange $s+\bar s$,  and  $s-\bar s$ (bottom
right),  all shown at $Q^2=2$~GeV. }
\end{figure}

In Fig.~\ref{fig:distances-HTp1} we show the distances between the PDFs determined including the higher twist correction
Eq.~(\ref{eq:FHT}) with $p_{\rm HT}=1$ and the default fit. The PDFs are mostly indistinguishable, the distance being 
compatible with statistical fluctuations. The only PDF which changes by a statistically significant amount is $s-\bar s$ 
(which is arguably the worst determined PDF combination in the context of a global fit), which undergoes a shift by about half 
sigma in the valence region. The other PDFs which are most affected, namely the total strangeness, valence and singlet change 
even less. The PDFs which change most are compared in Fig.~\ref{fig:comp-HTp1}: the changes are barely visible.

\begin{table}[t]
\centering
\footnotesize
 \begin{tabular}{c|c|c|c|c|c}
 \hline
 Exp & $N_{\rm dat}$ & $\chi^2$ (ref) & $\chi^2$ ($p_{\rm HT}$=1) & 
$\chi^2$ ($p_{\rm HT}$=2) & $\chi^2$ ($p_{\rm HT}$=-1) \\
 \hline
 \hline
Total &   3561    & 3989.4  &  4003.3   &   4096.6       &   4085.7             \\ 
\hline
NMC $d/p$         & 132     &        125.5  &  125.2        &    127.3      &   126.4        \\ 
NMC $p$           & 224      &        365.5  &  378.1        &   385.8       &  363.1         \\ 
SLAC           & 74     &         74.7  &     51.7          &    50.9      &   124.6        \\ 
BCDMS          & 581     &        768.6  &  790.9            &   892.4       &  811.2         \\ 
HERA-I      &  592    &        611.9  &     610.7            &   609.7       &  618.3         \\ 
CHORUS         &  862    &        959.9  &  957.6            &   956.6       &  959.5         \\ 
H1 $F_L$         &   8   &          9.8  &   9.8           &    9.8      &    9.8       \\ 
NuTeV         &  79    &         44.2  &   47.0            &    44.3      &    41.5       \\ 
ZEUS HERA-II    & 127    &        163.5  &   164.0          &    163.8      &   163.3        \\ 
ZEUS $F_2^c$     & 62   &         60.9  &    60.2             &    60.9      &   62.3        \\ 
H1 $F_2^c$       & 45    &         69.4  &   69.2              &    68.9      &   70.1         \\ 
DY E605          & 119    &         98.6  &  100.4                   & 97.8         &  98.0          \\ 
DY E886         &  199    &        263.2  &  258.0                   &  253.7        & 260.9          \\ 
CDF $W$ asy     &  13    &         21.2  &   22.0                 &  21.6        &      21.4     \\ 
CDF $Z$ rap     &  29   &         53.2  &    57.7         &       62.7   &        52.2   \\ 
D0 $Z$ rap      &  28   &         17.5  &    17.7            &     18.2     &      17.6     \\ 
ATLAS $W,Z$     &   30  &         41.4  &    43.1                &   42.9       &  41.4         \\ 
CMS $W$ el asy   &  11  &          8.9  &    9.5                &     9.4     &    8.6       \\ 
LHCb $W,Z$       &  10  &          7.8  &    7.8                &     7.8     &    7.6       \\ 
CDF RII $k_T$    &  76   &         59.6  &   59.77                 &   53.9       &   59.6        \\ 
D0 RII cone     &   110  &         90.6  &   89.7                 &   87.2       &   93.5        \\ 
ATLAS jets      &   90  &         73.3  &    72.7               &    70.7      &    74.7       \\ 
 \hline
 \end{tabular}
\caption{\small The $\chi^2$ of the global fit before and after the
  inclusion of higher twist corrections, for the three
scenarios shown in Fig.~\ref{fig:ht_corr}. We also provide the number
of data points for each experiment. Note that this is the absolute
$\chi^2$, not divided by the number of data points.
 \label{tab:httab} }
\end{table}

\begin{figure}[h!]
    \begin{center}
\includegraphics[width=0.90\textwidth]{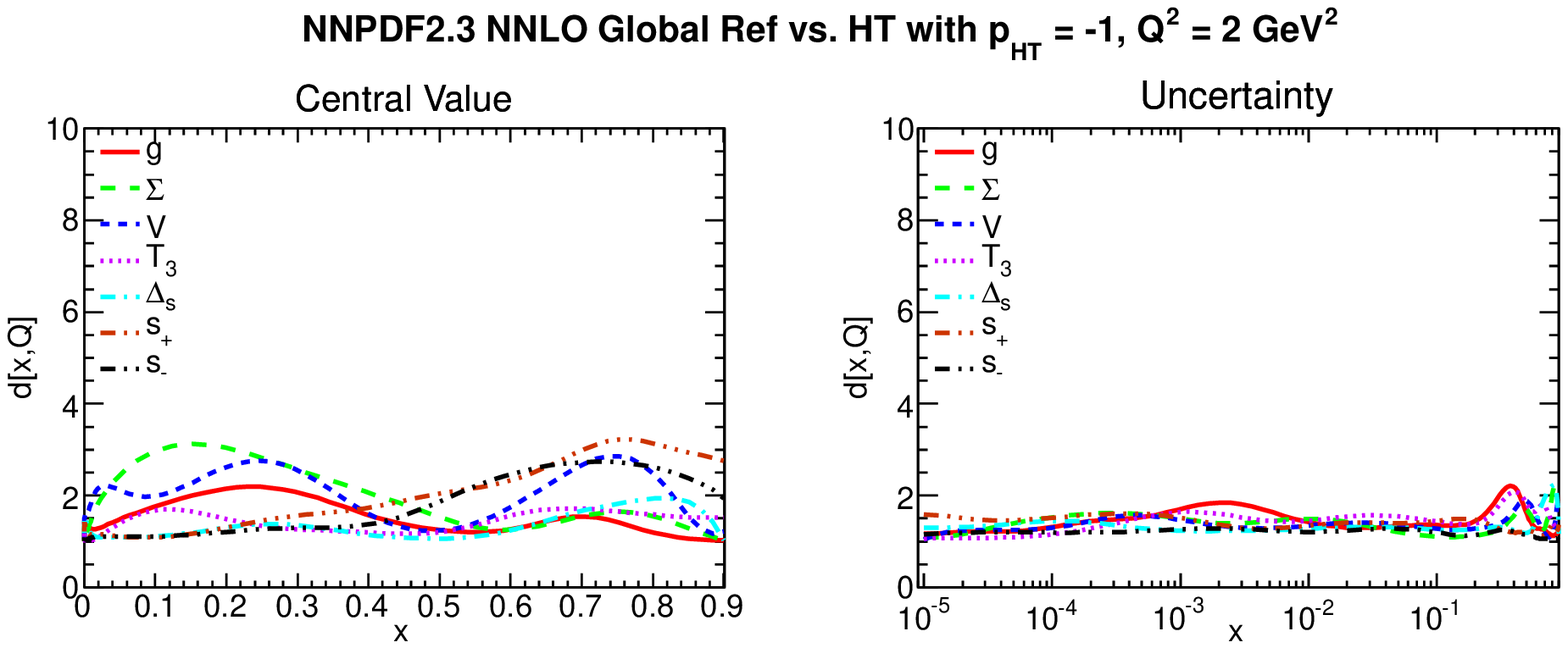}
\includegraphics[width=0.90\textwidth]{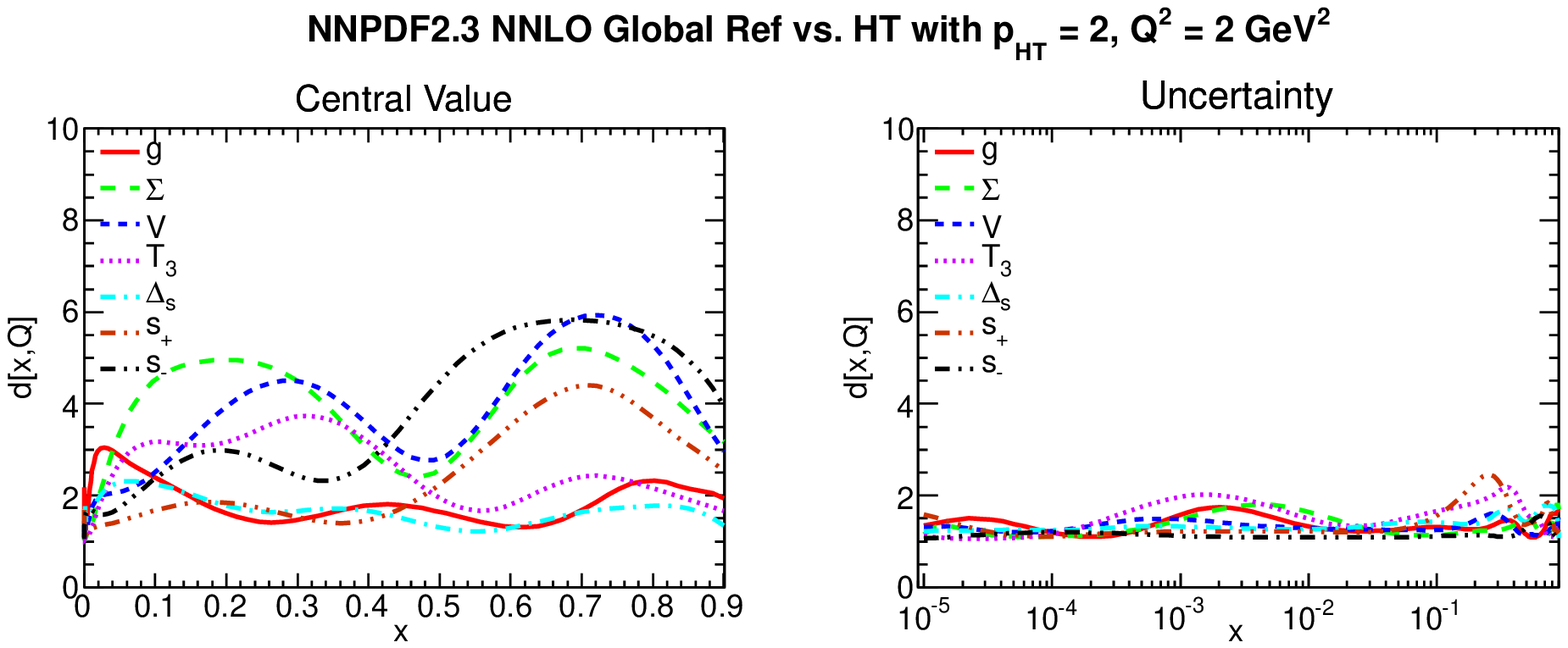}
      \end{center}
     \caption{\small 
    \label{fig:distances-HTp2} 
Same as Fig.~\ref{fig:distances-HTp1}, but with 
$p_{\rm HT}=-1$ (top row) and $p_{\rm HT}=-2$ (bottom row).}
\end{figure}

We have then repeated the PDF determination with the extreme choices $p_{\rm HT}=-1$ and $p_{HT}=2$ in Eq.~(\ref{eq:FHT}). 
Distances are shown in Fig.~\ref{fig:distances-HTp2}. It is clear that when the sign of higher twist corrections is reversed, their 
effect remains negligible, and even when they are arbitrarily doubled, PDFs always change by less than half sigma, and mostly 
much less than that. This fit is performed with the same
default cut  $W^2 \ge 12.5$ GeV$^2$ adopted in NNPDF PDF
determinations: it appears that with this cut,
the impact of including higher twist corrections to DIS structure
functions is negligible. The same conclusion is very likely to apply
to the MSTW08 and CT10 PDF determinations, which adopt similar cuts.

In Table~\ref{tab:httab} we show the $\chi^2$ for the three fits
including higher twist corrections, both for the global dataset, and
for each experiment. It appears that when $p_{\rm HT}=1$ the global
$\chi^2$ is essentially unchanged, with the improvement in the
desription of the SLAC data compensated by the deterioration of
the fit to BCDMS data. With the other two choices, $p_{\rm HT}=-1$ or
$p_{\rm HT}=2$, the fit quality deteriorates substantially, as one
might expect.

\begin{figure}[h!]
    \begin{center}
\includegraphics[width=0.90\textwidth]{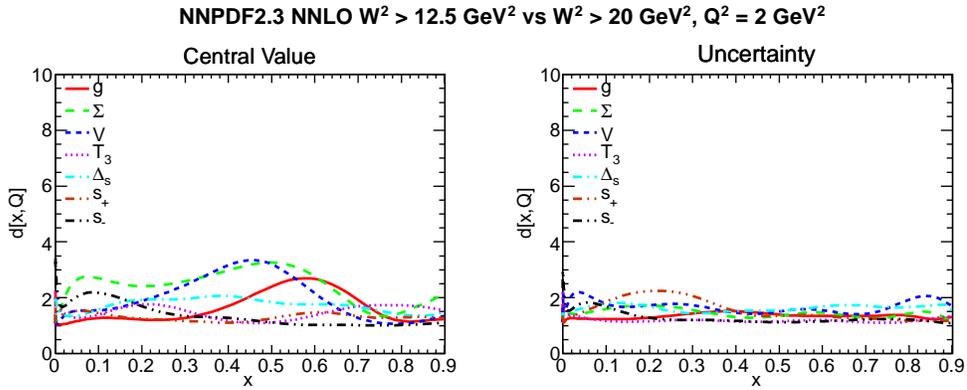}
      \end{center}
     \caption{\small 
    \label{fig:distances-higherW2} 
Distances between the reference NNPDF2.3 PDFs, with $W^2\ge 12.5$ GeV$^2$, 
and PDFs obtained imposing on the dataset the  tighter constraint
$W^2\ge 20$ GeV$^2$.} 
\end{figure}

As a final consistency check, we have repeated the standard NNPDF2.3 PDF determination with no higher twist
corrections, but with a stricter cut  $W^2 \ge 20$ GeV$^2$. With this cut, any  possible residual effect of higher twists in the 
default fit would be greatly reduced, and thus the variation of results is an indication of their possible presence in the default 
fit. The distances between this PDF set and the default are shown in Fig.~\ref{fig:distances-higherW2}. Again, they are barely 
above the level of statistical fluctuations --- because the more stringent cut changes the dataset, full statistical equivalence is 
not expected, but changes are below, usually much below, the half sigma level, and thus not statistically significant.
We conclude that higher twist corrections and associated uncertainties
are negligible in current global NNPDF sets, and so is their impact on
the quality of the global fit.
Similar results were found in a related MSTW analysis~\cite{Thorne:2011kq}.

\bigskip\noindent
{\bf Nuclear Corrections}

Finally, we discuss the impact of nuclear corrections on PDF determinations. In the NNPDF2.3 fit (and in other global fits such 
as MSTW08 and CT10) three classes of data which may be affected by nuclear corrections are used: neutrino DIS data, which
are obtained on heavy, approximately isoscalar, nuclear targets (such as iron); fixed-target data for DIS on deuterium, 
and fixed-target Drell-Yan data on deuterium. The impact of nuclear corrections on neutrino DIS data was studied in 
Ref.~\cite{Ball:2009mk}, and found to be negligible in comparison to the sizable uncertainties on these data. Nuclear corrections 
to deuterium are rather smaller than those for heavy nuclei, but structure function data on deuterium targets can be quite precise, so here 
they could have an impact, especially on the determination of the up-down quark ratio at large $x$.

The possible impact of deuterium nuclear corrections was recently emphasized in Ref.~\cite{Owens:2012bv} where, relying on 
previous studies~\cite{Accardi:2009br,Accardi:2011fa},  the CJ12 PDF sets were presented, based on CTEQ methodology but 
including nuclear corrections to deuterium structure function data derived using a variety of models. The impact of deuterium
corrections was also recently studied in Ref.~\cite{Martin:2012xx}, where they were fitted to the data.

\begin{figure}[h!]
    \begin{center}
\includegraphics[width=0.60\textwidth]{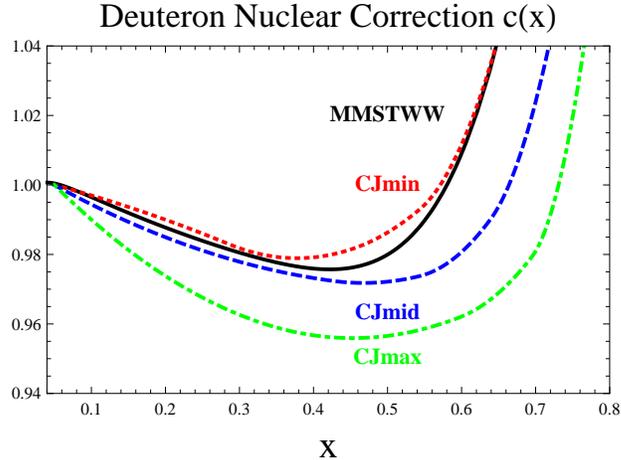}
      \end{center}
     \caption{\small 
    \label{fig:deutcorr} 
The nuclear correction factor
Eq.~(\ref{nuclcorr}) from the fit of Ref.~\cite{Martin:2012xx} (labeled MMSTWW)
and from the three models of Ref.~\cite{Owens:2012bv} (labeled CJmin, CJmid
 and CJmax). The  MMSTWW correction
factor is $Q^2$ independent, while the CJ models
are shown  for  $Q^2=100$~GeV$^2$.
 }
\end{figure}

We have studied the impact of deuterium corrections on the NNPDF2.3 PDF determination by correcting all deuterium structure 
function data according to 
\be\label{nuclcorr}
F_2^d(x,Q^2)=c(x)\lp F_2^p(x,Q^2)+F_2^n(x,Q^2)\rp/2 \, .
\ee
For the correction factor $c(x)$ we have first adopted the phenomenological determination obtained in Ref.~\cite{Martin:2012xx}.
This is $Q^2$-independent and such that the correction is negative below a given value of $x$ and positive above it, and is 
parametrized by three parameters determined through a global PDF fit based on the MSTW08 methodology. 
We have also computed  $c(x)$ for the three choices considered in Ref.~\cite{Owens:2012bv} (CJmin, CJmid and CJmax) using the 
expressions of $F_2^d(x,Q^2)$, $F_2^p(x,Q^2)$ and $F_2^n(x,Q^2)$
provided by the authors. 

Eq.~(\ref{nuclcorr}) should be viewed as a $K$-factor approximation, because in the nuclear models used in
Ref.~\cite{Owens:2012bv} the nuclear correction is not just  multiplicative, but rather it is a $Q^2$-dependent correction which 
depends on the structure function itself, partly in a convolutive way. This approximation is adequate for our current goal, which
is to determine the size of these corrections and their associated
uncertainties, rather than their shape.  Even though the correction of
Ref.~\cite{Owens:2012bv} is scale dependent, we have evaluated it at 
$Q^2=100$~GeV$^2$
and we have assumed it to be $Q^2$-independent, as the $Q^2$
dependence is weak in the region $x\lsim 0.5$~\cite{Accardi:2011bc} 
where, as we shall see,
the impact of the correction is significant.
The shape of the nuclear correction Eq.~(\ref{nuclcorr}) in all these cases is displayed in Fig.~\ref{fig:deutcorr}.

\begin{figure}[h!]
    \begin{center}
\includegraphics[width=0.48\textwidth]{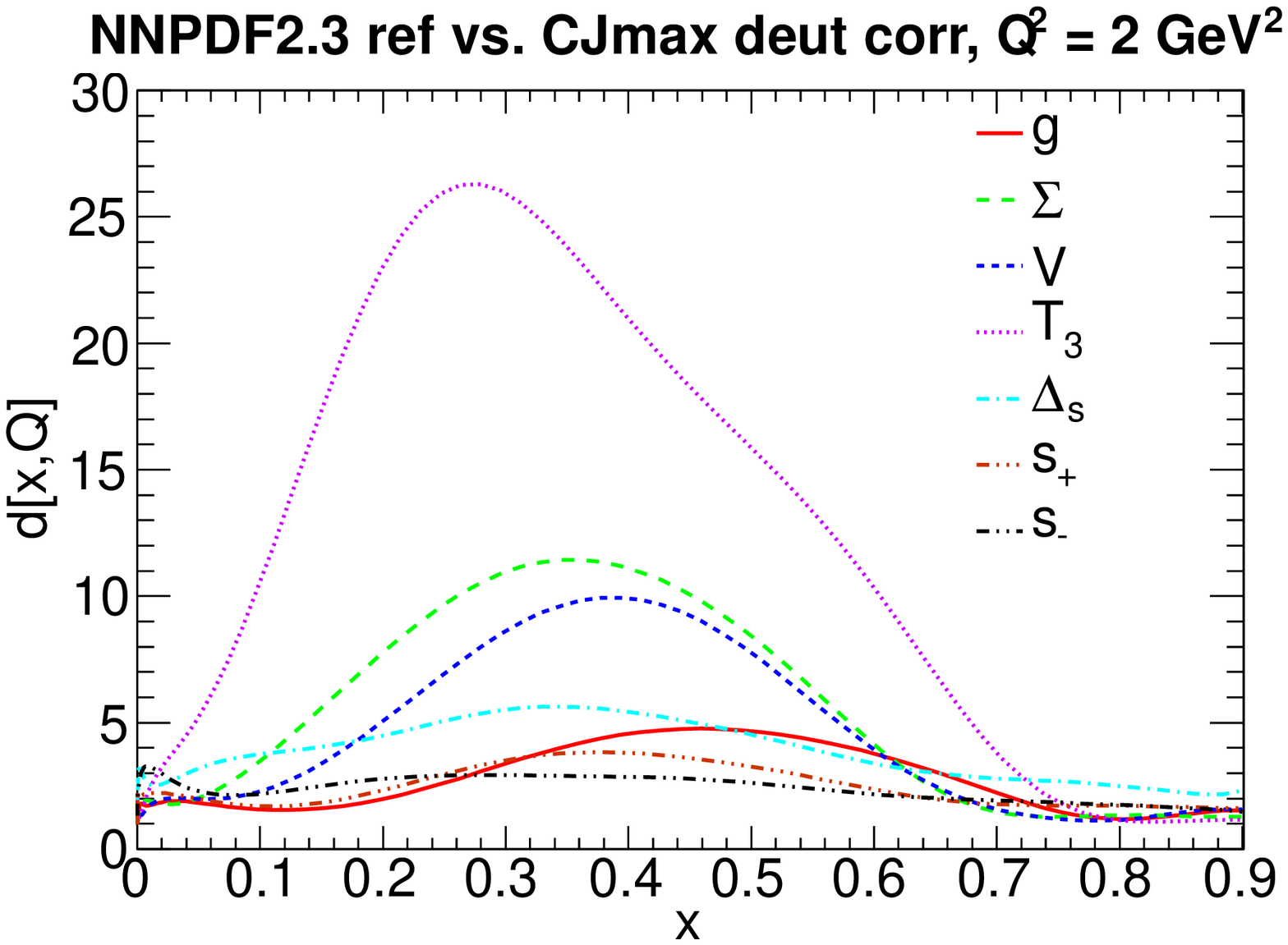}
\includegraphics[width=0.48\textwidth]{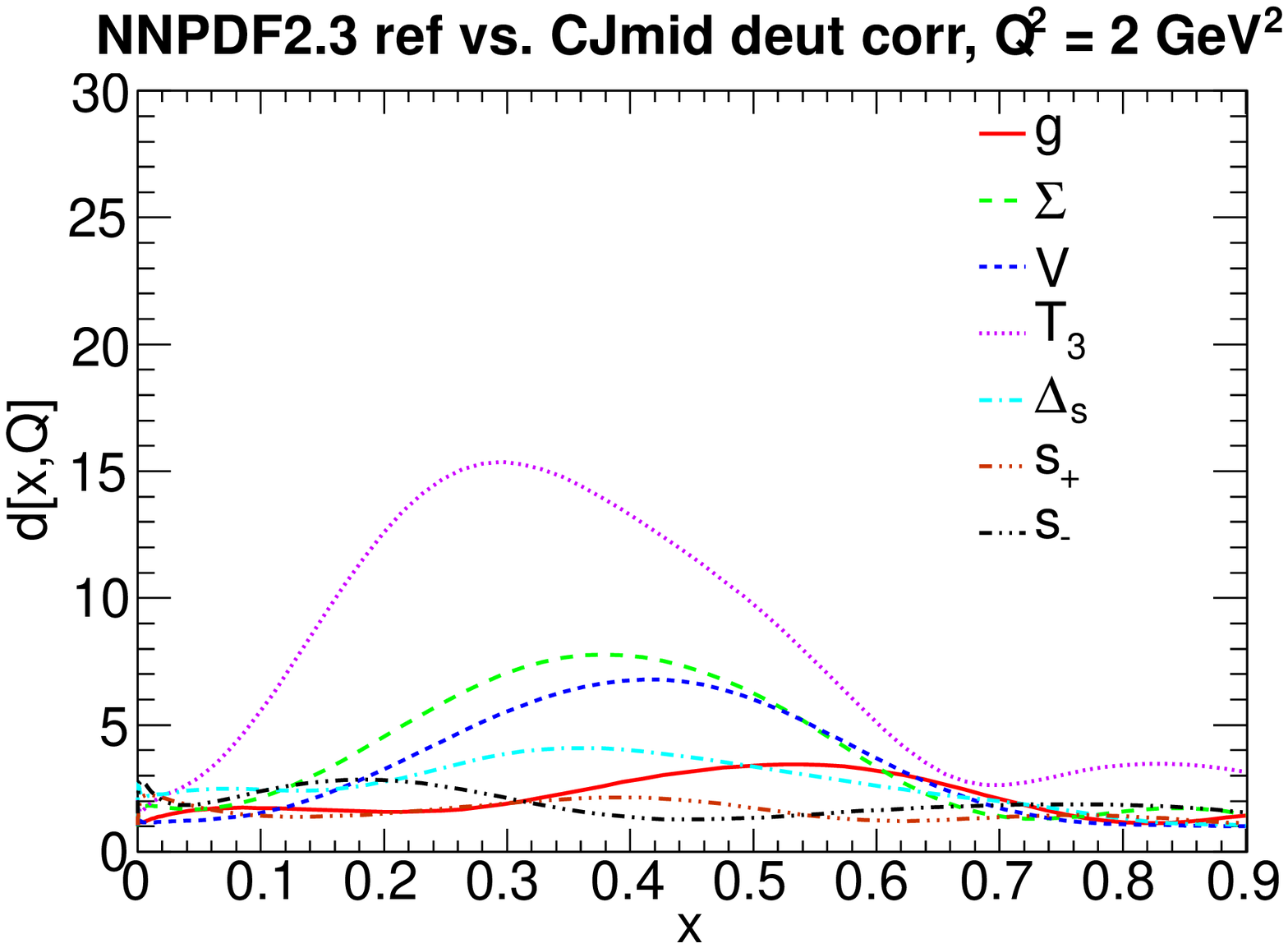}
\includegraphics[width=0.48\textwidth]{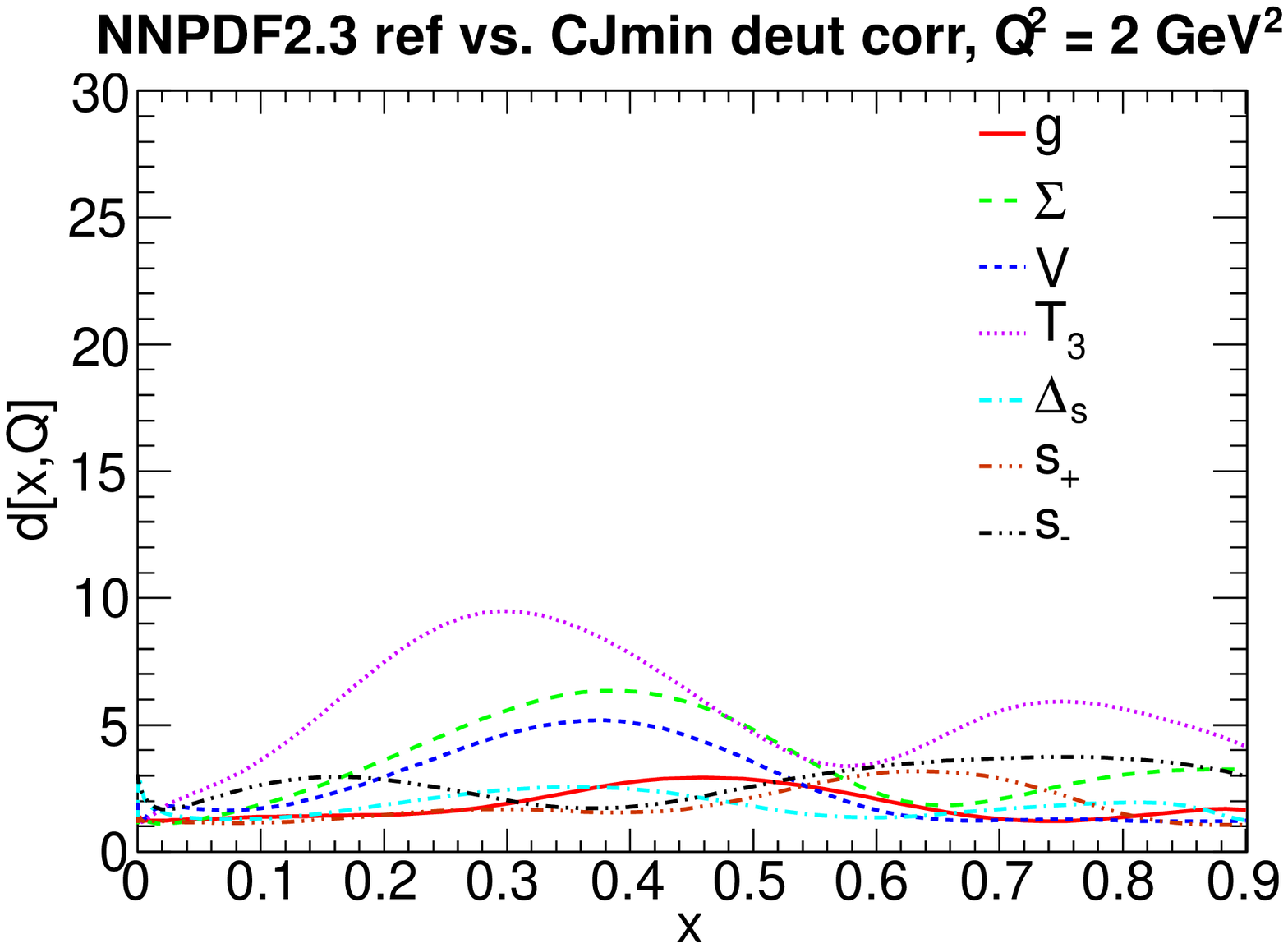}
\includegraphics[width=0.48\textwidth]{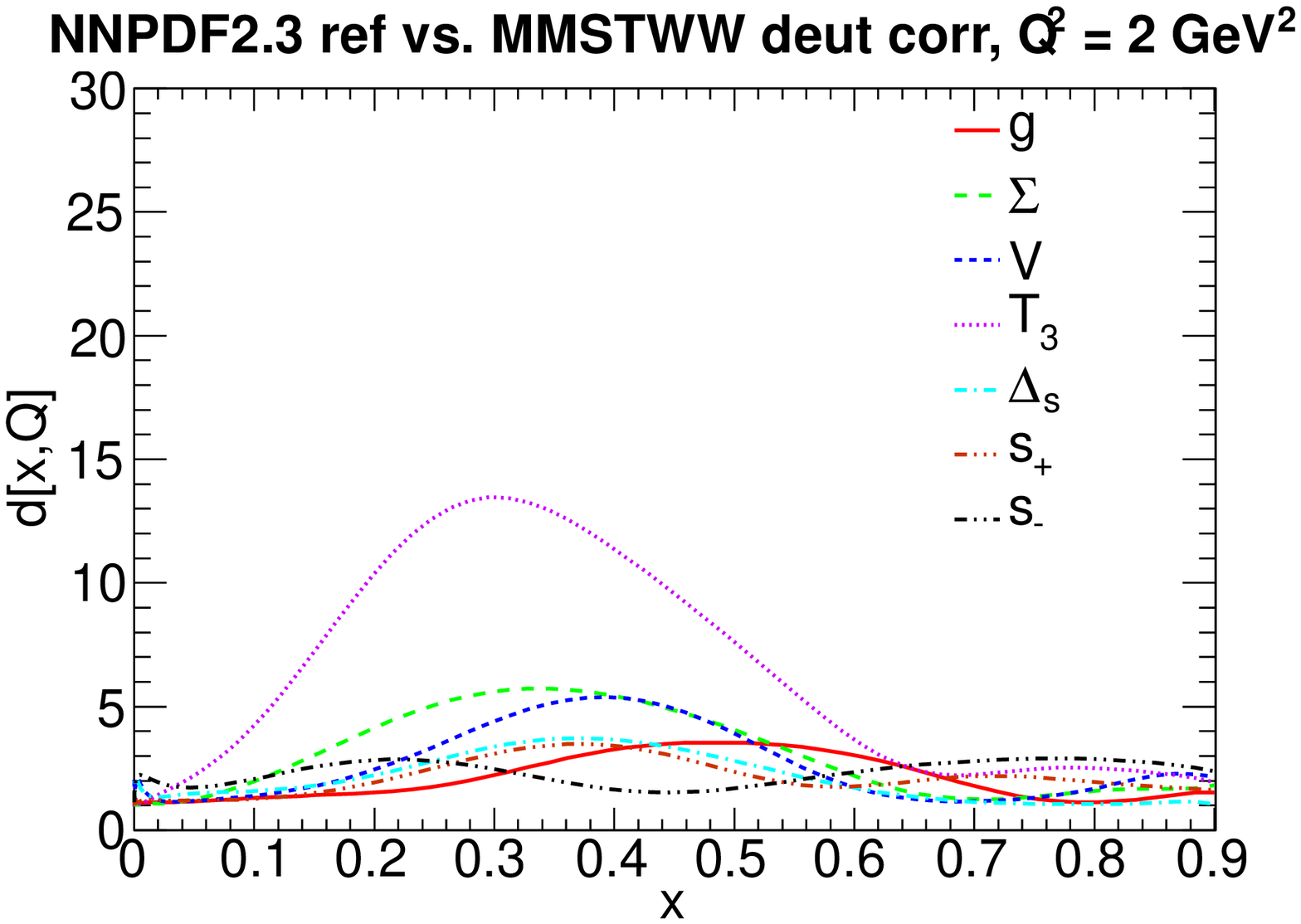}
      \end{center}
     \caption{\small 
    \label{fig:distances-deuteron} 
Distances between the reference NNPDF2.3 PDFs, and PDFs obtained by
introducing nuclear corrections to deuterium structure function data
according to Eq.~(\ref{nuclcorr}), with the four nuclear correction factors shown
in Fig.~\ref{fig:deutcorr}. Only distances between central values are shown.
}
\end{figure}

\begin{figure}[h!]
    \begin{center}
\includegraphics[width=0.45\textwidth]{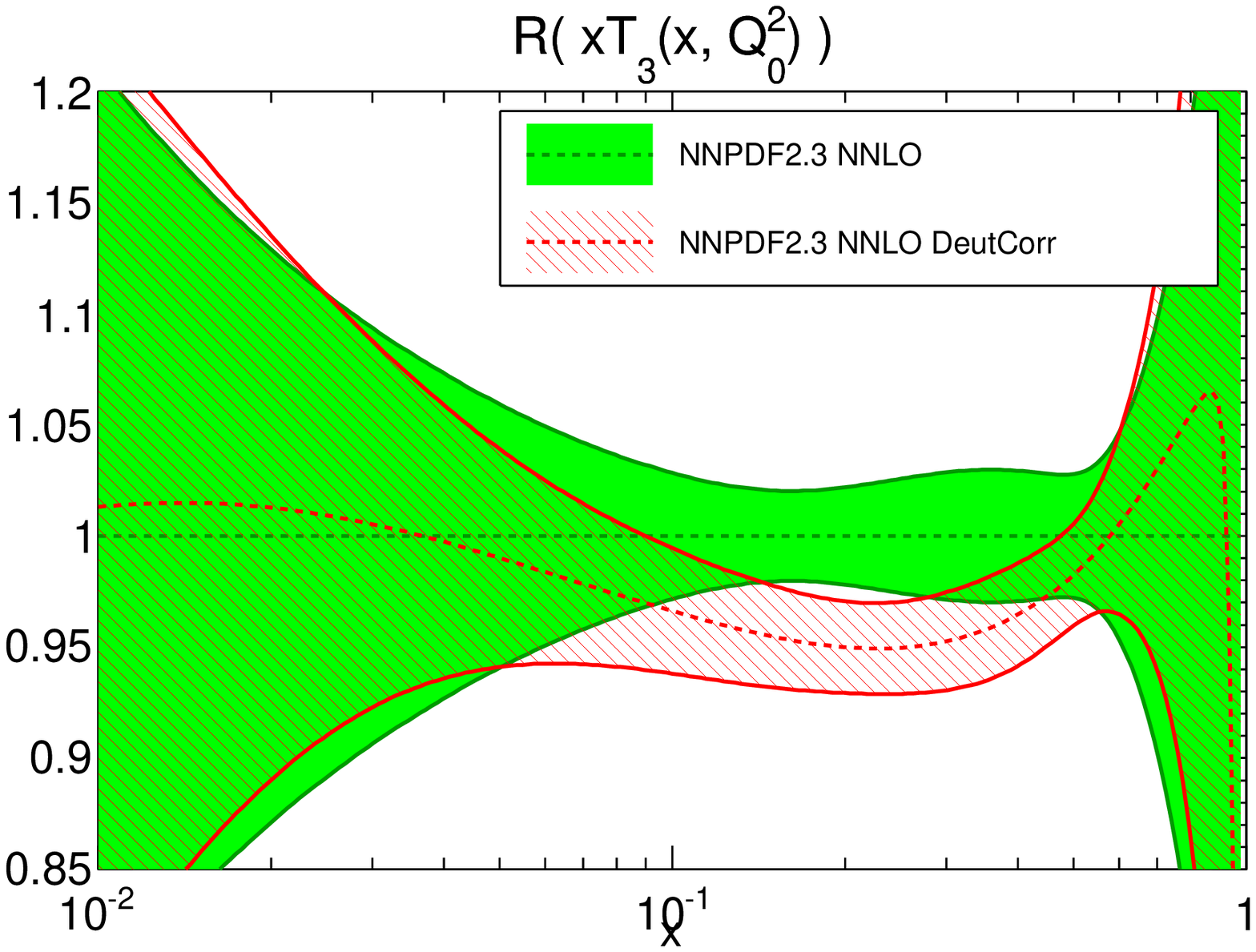}
\includegraphics[width=0.45\textwidth]{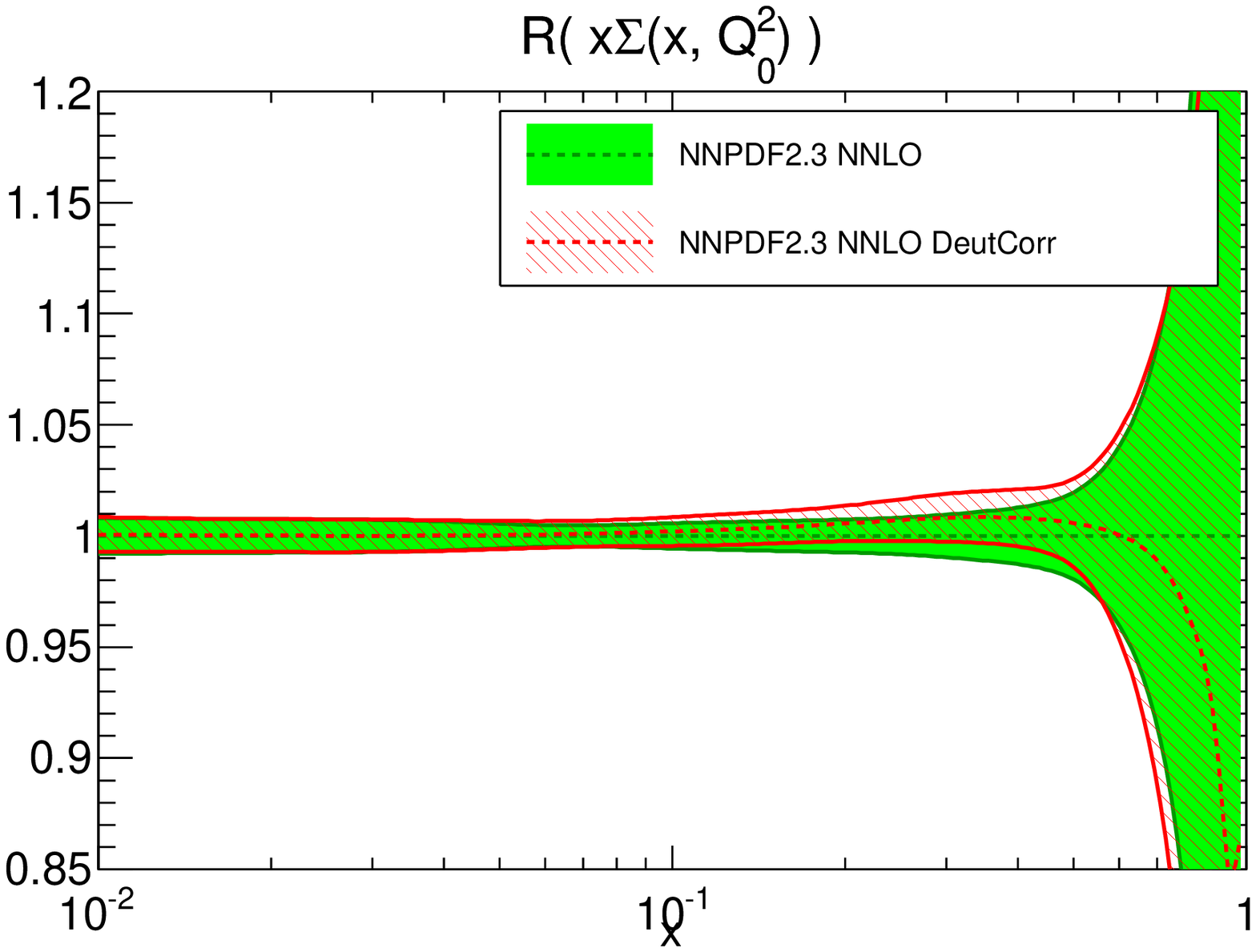}
      \end{center}
     \caption{\small 
    \label{fig:pdf-rat-deut} 
Comparison of PDFs which are mostly affected by deuterium nuclear corrections,
using the MMSTWW model,
 and the default (distances were shown   in
Fig.~\ref{fig:distances-deuteron}): isotriplet (left) and singlet
(right), 
all shown at
     $Q^2=2$~GeV$^2$.}
\end{figure}

We have then repeated the NNPDF2.3 PDF determination including the correction according to each of these four models in turn. 
Note that only deuterium structure function data are corrected, so in particular Drell-Yan data on fixed deuterium targets remain 
uncorrected: indeed, Refs.~\cite{Martin:2012xx,Owens:2012bv} only
consider deuterium corrections to DIS structure functions. These data,
however, are mostly in a kinematic region where nuclear corrections are very
small.
The distances between PDFs obtained  including nuclear corrections in this way, and the default NNPDF2.3 PDFs, are displayed in
Fig.~\ref{fig:distances-deuteron}. We only show distances between central values, as we have verified that
uncertainties are unaffected. The only PDF combination which is significantly affected by the introduction of deuterium corrections
is the isospin triplet, which changes by more than one sigma for $0.1\lsim x\lsim0.5$  and up to one and a half sigma at 
the valence peak $x\sim0.3$, for the intermediate MMSTWW and CJmid cases. Changes of up to half sigma at the peak are seen
also in the valence and $\Delta_s=\bar d-\bar u$ combinations, though these, as well as all other PDFs, mostly display changes 
which are compatible with statistical fluctuations.
In the extreme CJmax case, the variation can be up to three sigma at the peak for the isospin triplet, and up to one sigma for the 
singlet and $\Delta_s$. 
A comparison between the default fit, and that with nuclear corrections included following Ref.~\cite{Martin:2012xx}, is presented 
in Fig.~\ref{fig:pdf-rat-deut} for the two PDF combinations which change most, namely the  triplet and singlet.
 
\begin{table}[t]
\centering
\footnotesize
 \begin{tabular}{c|c|c|c|c|c|c}
 \hline
 Exp & $N_{\rm dat}$ & $\chi^2$ (ref) & $\chi^2$ (+MMSTWW) & 
$\chi^2$ (+CJmin) & $\chi^2$ (+CJmid) & $\chi^2$ (+Cjmax)\\ 
 \hline
 \hline
Total &   3561    & 3989.4  &        4024.1  &        4051.0  &        4039.1  &        4102.1  \\ 
\hline
NMC $d/p$         & 132     &        125.5  &         126.6  &         130.8  &         126.5  &         132.7  \\ 
NMC $p$           & 224      &        365.5  &         359.2  &         363.7  &         360.0  &         367.7  \\ 
SLAC           & 74     &         74.7  &          85.5  &          80.1  &          82.5  &          93.2  \\ 
BCDMS          & 581     &        768.6  &         775.5  &         788.6  &         772.9  &         795.4  \\ 
HERA-I      &  592    &        611.9  &         609.7  &         618.1  &         611.5  &         609.9  \\ 
CHORUS         &  862    &        959.9  &         977.5  &         981.7  &         989.6  &         996.4  \\ 
H1 $F_L$         &   8   &          9.8  &           9.8  &          10.2  &           9.8  &           9.9  \\ 
NuTeV         &  79    &         44.2  &          49.5  &          44.6  &          52.0  &          56.2  \\ 
ZEUS HERA-II    & 127    &        163.5  &         163.8  &         162.7  &         162.9  &         163.3  \\ 
ZEUS $F_2^c$     & 62   &         60.9  &          60.8  &          58.0  &          61.0  &          58.7  \\ 
H1 $F_2^c$       & 45    &         69.4  &          69.2  &          68.0  &          68.6  &          67.9  \\ 
DY E605          & 119    &         98.6  &         100.7  &         100.1  &         103.0  &         102.9  \\ 
DY E886         &  199    &        263.2  &         258.7  &         267.4  &         265.0  &         265.4  \\ 
CDF $W$ asy     &  13    &         21.2  &          18.3  &          22.5  &          18.9  &          19.1  \\ 
CDF $Z$ rap     &  29   &         53.2  &          57.9  &          58.1  &          57.8  &          60.5  \\ 
D0 $Z$ rap      &  28   &         17.5  &          18.2  &          18.3  &          18.3  &          18.8  \\ 
ATLAS $W,Z$     &   30  &         41.4  &          41.8  &          43.6  &          41.5  &          41.7  \\ 
CMS $W$ el asy   &  11  &          8.9  &           9.0  &           9.0  &           8.4  &           8.7  \\ 
LHCb $W,Z$       &  10  &          7.8  &           7.5  &           7.3  &           7.5  &           8.7  \\ 
CDF RII $k_T$    &  76   &         59.6  &          61.4  &          57.5  &          58.4  &          62.0  \\ 
D0 RII cone     &   110  &         90.6  &          90.5  &          89.6  &          90.4  &          90.0  \\ 
ATLAS jets      &   90  &         73.3  &          73.1  &          71.1  &          72.7  &          72.9  \\ 
 \hline
 \end{tabular}


\caption{\small The $\chi^2$ of the global fit before and after the
  inclusion of deuterium nuclear corrections, with the four models
  shown in Fig.~\ref{fig:deutcorr}. We also provide the number
of data points for each experiment. Note that this is the absolute
$\chi^2$, not divided by the number of data points.
 \label{tab:deuttab} }
\end{table}

In Ref.~\cite{Owens:2012bv} it was argued that the quality of the global fit is essentially unaffected by these nuclear corrections,
which are absorbed in a change of the PDFs, but only if they are not too large, and thus in particular it was argued that deuterium
corrections as large as  CJmax are disfavored by the data. In
Ref.~\cite{Martin:2012xx} a moderate but non-negligble improvement in
fit quality was
found when the nuclear corrections are added.
In our case, we find that the  fit quality is essentially unaffected
by the inclusion of nuclear correction, unless they are too large, in
which case the fit quality deteriorates significantly. 
In Table~\ref{tab:deuttab} 
we show the $\chi^2$  for the global fit and for each dataset included in it, both for the default fit and for the fits with deuterium corrections. 
The fit quality deteriorates somewhat upon inclusion of nuclear
corrections, by an amount per datapoint ($\Delta \chi^2\approx0.01$) 
which (with the MMSTWW form of the correction) is about half the
improvement seen in Ref.~\cite{Martin:2012xx}. It is interesting to
observe that most of this deterioration comes from the CHORUS neutrino
deep-inelastic scattering data, which are obtained using heavy nuclear
targets. These data are corrected for nuclear effects in
Ref.~\cite{Martin:2012xx}, but not in our study, which might explain
the difference. When  CJmin and CJmid corrections are applied, the fit
quality deteriorates by a smilar or somewhat larger amount (also
mostly due to CHORUS data), while it deteriorates significantly if 
CJmax corrections are used, in agreement with the findings of
Ref.~\cite{Owens:2012bv}. 

\begin{figure}[h!]
    \begin{center}
\includegraphics[width=0.48\textwidth]{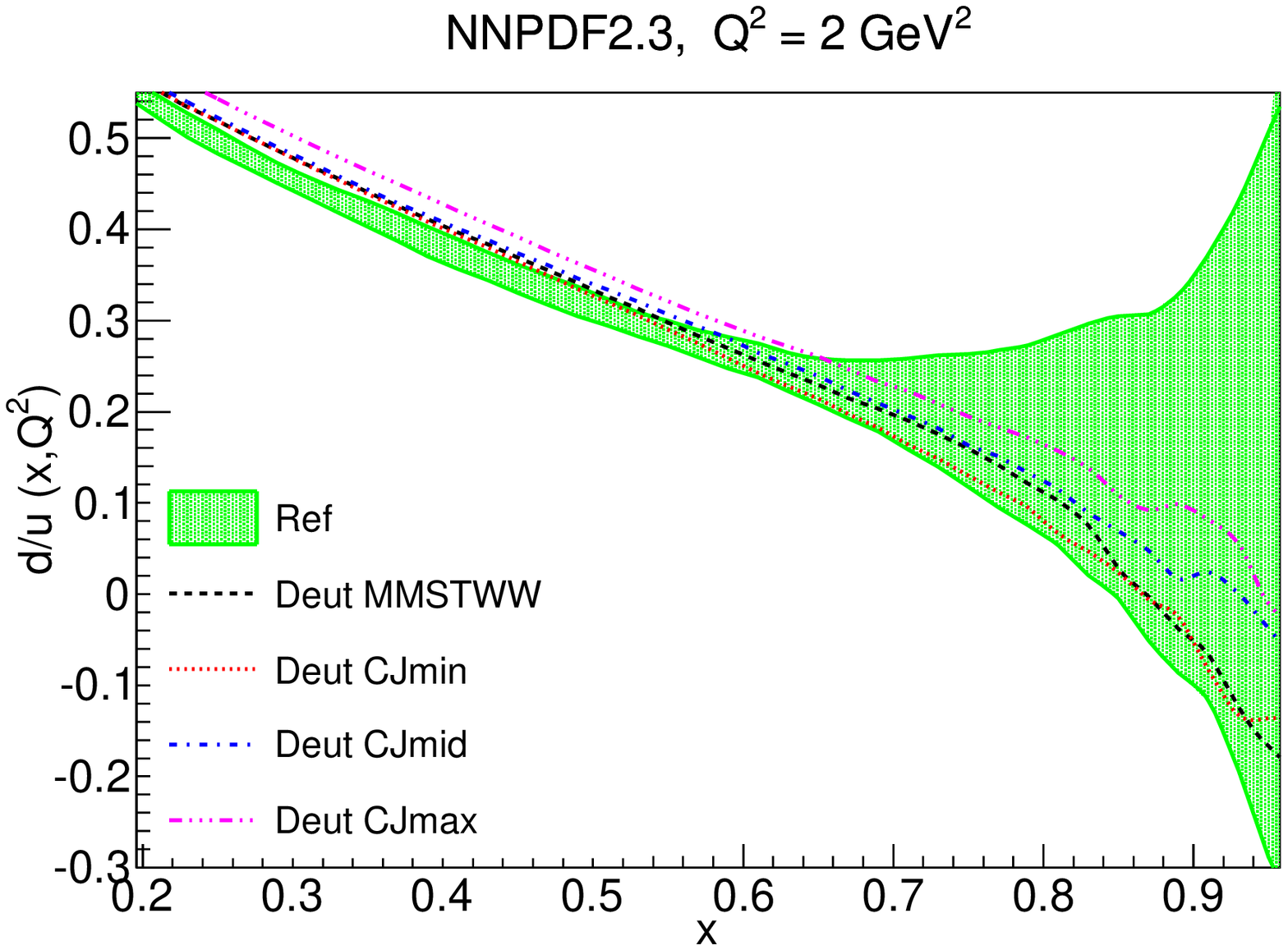}
\includegraphics[width=0.48\textwidth]{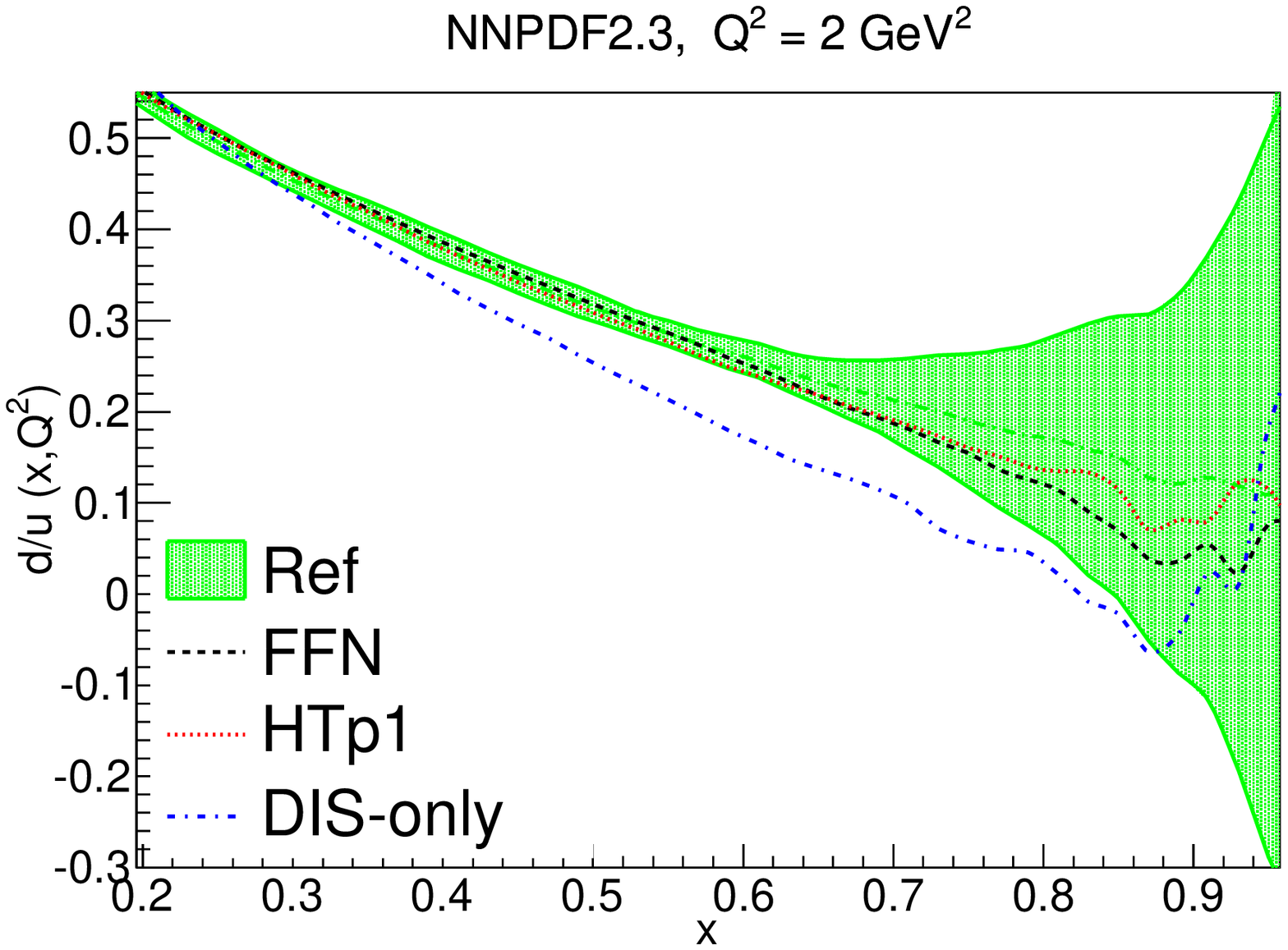}
      \end{center}
     \caption{\small 
    \label{fig:d_over_u} 
The $d/u$ ratio at $Q^2=2$ GeV$^2$ as a function of $x$. The 
central value obtained with several variants of the fit is compared to
the default NNPDF2.3 result and uncertainty. Left: effect of different
models of deuterium corrections shown in
Fig.~\ref{fig:deutcorr}. Right: effect of
using a FFN scheme, of using DIS-only data,
 and of including the higher twist correction 
shown in Fig.~\ref{fig:ht_corr} with the central choice $p_{\rm HT}=1$.
PDF uncertainties are computed as 68\% confidence levels.}
\end{figure}

The role of nuclear corrections in determining the down/up ratio in the  $x\to1$ limit has been especially emphasized in
Refs.~\cite{Accardi:2009br,Accardi:2011fa,Owens:2012bv}. In order to elucidate this point, in Fig.~\ref{fig:d_over_u} we look at 
the $d(x)/u(x)$ ratio  at $Q^2=2$~GeV$^2$: we show the NNPDF2.3 result with uncertainty, and we superimpose on it the central value 
of the ratio obtained in fits where different types of deuterium nuclear corrections are included (left plot). 
For comparison, we also show the effect on the central value of the ratio of only fitting to DIS data,  of using a FFN scheme, 
and of including higher twists according to Eq.~(\ref{eq:FHT}) with the default choice $p_{\rm HT}=1$ (right plot). PDF uncertainties
are computed as 68\% CL, since  at large-$x$ uncertainties show
non-gaussian behaviour.
It is clear that the impact of deuterium corrections is visible for $x\lsim 0.5$, as already seen in the distances plots of 
Fig.~\ref{fig:distances-deuteron}, but it is completely negligible in comparison to the uncertainty for larger $x$ values. Similarly 
negligible are the impact of higher twist corrections, and even of the use of a FFN scheme. By contrast, what does have a significant impact, 
up to very large $x\sim0.8$ is the use of DIS only data in the PDF
determination. 

We can therefore conclude that the use of a global dataset, including in particular a wide variety of hadronic data, is crucial in order 
to have a handle on the large $x$ flavor separation. On the other hand for $x\gsim 0.5$, the light flavor separation is affected by a large
uncertainty due to the scarcity of experimental information. Theoretical uncertainties related to the effects which we study here, and 
in particular nuclear corrections are completely negligible on the
scale of these uncertainties. A somewhat different conclusion was
reached in
Ref.~\cite{Owens:2012bv}, where it was argued that nuclear corrections
affect signficantly the  $d(x)/u(x)$ ratio in the $x\to1$ limit. 
In this reference, a wider dataset was used,
including low $Q^2$ and low $W^2$ data, while higher twist and nuclear
corrections were introduced. Whereas inclusion of such data might raise somewhat
the value of $x$ at which uncertainties start blowing up, it appears
that when using a the more general NNPDF parametrization, rather than
the more restrictive one used in  Ref.~\cite{Owens:2012bv}, the
uncertainties on the $d(x)/u(x)$ ratio in the $x\to1$ limit are
necessarily so large that nuclear corrections are unlikely to play a
significant role.

\begin{figure}[t]
    \begin{center}
\includegraphics[width=0.68\textwidth]{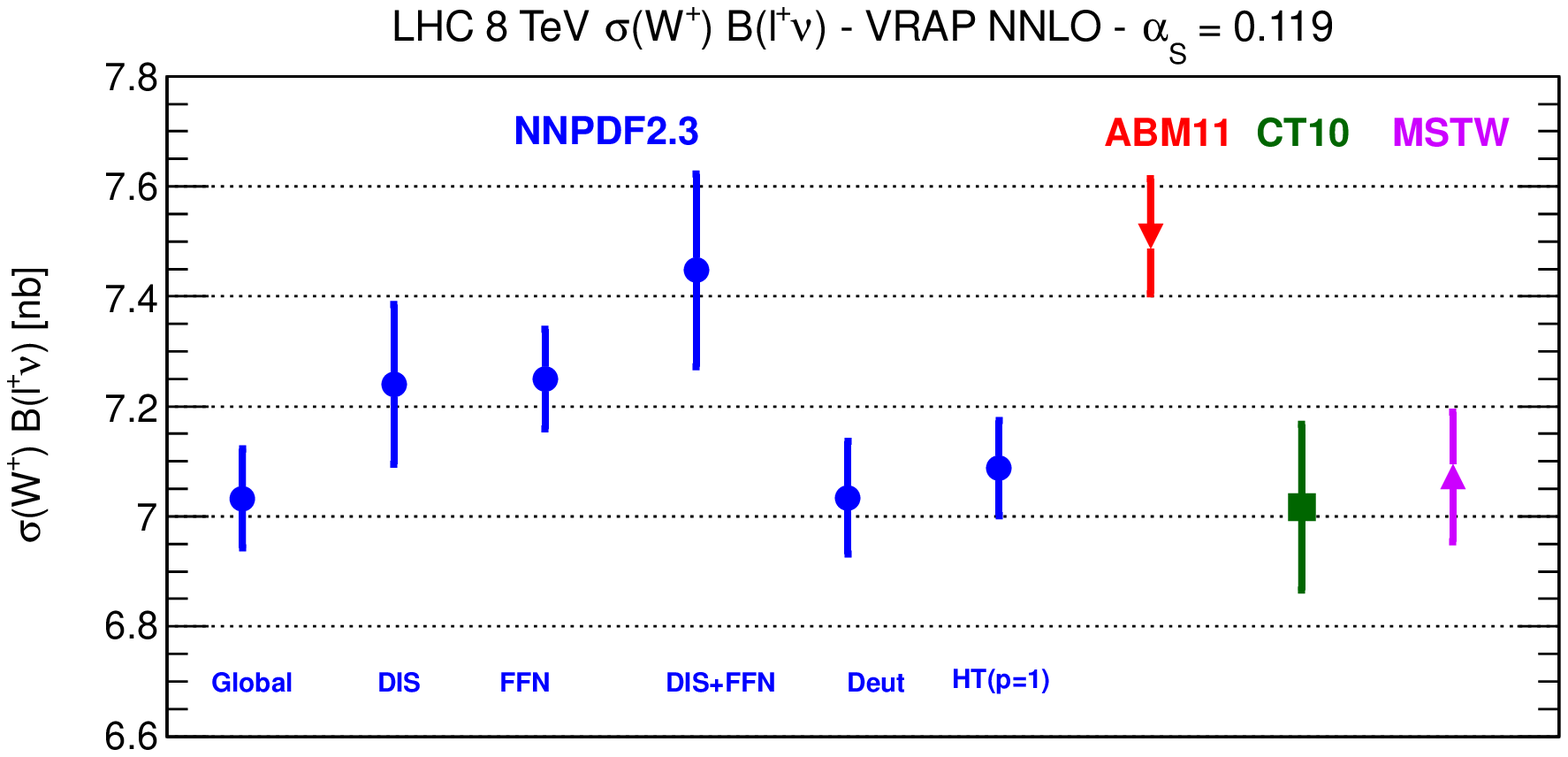}
\includegraphics[width=0.68\textwidth]{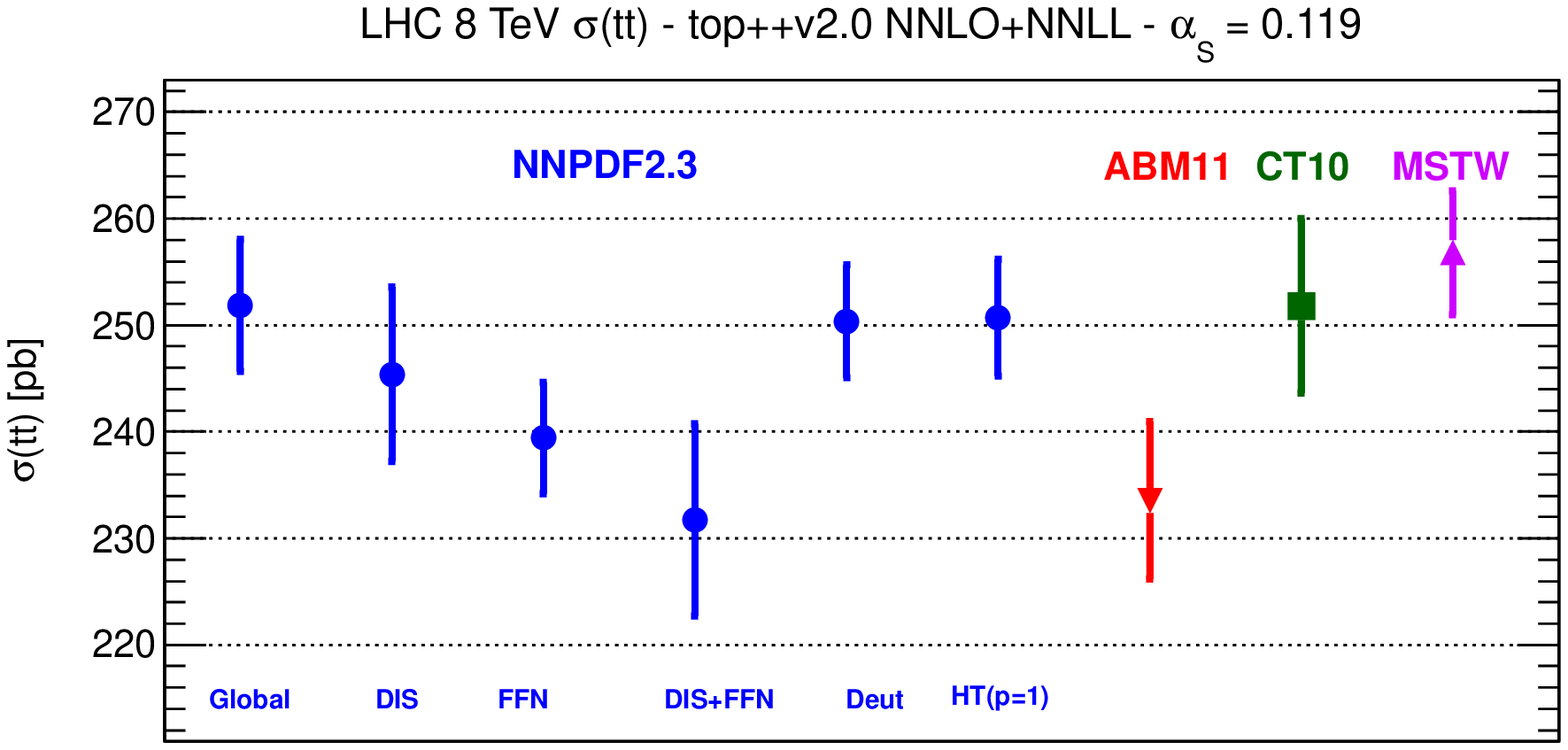}
      \end{center}
     \caption{\small 
    \label{fig:lhcxsec} Some LHC standard candles computed at NNLO using
    various PDF sets discussed in this paper:
$W^+$ production (top) and $t\bar{t}$ production
(bottom) $\sqrt{s}=8$~TeV. All results are shown for
$\alpha_s(M_Z)=0.119$.}
\end{figure}

We conclude that the impact of deuterium nuclear corrections is
non-negligible in  the region $0.1\lsim x\lsim 0.5$ on the isospin 
triplet combination, i.e. on the up-down separation.
However, the theoretical status of these corrections is not
entirely satisfactory: a variety of models is available, but they are
not clearly favored by the data, and may become  disfavored if 
the correction is large. Given the uncertainties involved, the inclusion
of such corrections is not clearly advantageous at present. However,  it 
should be 
kept in mind that the uncertainty on the isotriplet  should be supplemented 
by a theoretical uncertainty related to nuclear corrections in the
$0.1\lsim x\lsim 0.5$ region.  

\bigskip\noindent
{\bf Conclusions}
\nobreak

We summarize our results by looking at the impact of the three different sources of theoretical uncertainties considered here on the predictions for a representative set of LHC standard candles, namely the  total cross sections for $W^+$ production, 
which is sensitive to quark and antiquark distributions, and top production, which is sensitive to the gluon distribution at large 
Bjorken-$x$.
We have computed these processes at NNLO, using the default NNPDF2.3 PDFs and various other sets discussed in this paper, 
as well as with other PDF sets which we have referred to in the course of the present discussion, namely  ABM11~\cite{Alekhin:2012ig},
CT10~\cite{Nadolsky:2009ge,Nadolsky:2012ia}  and MSTW08~\cite{Martin:2009iq}, all with a common value of $\alpha_s(M_z)=0.119$. 
The codes and settings used to compute the various cross sections are
the same as in the recent benchmark study Ref.~\cite{Ball:2012wy}; for
top production we have used the more recent version 2.0 of the {\tt
  Top++} code~\cite{Czakon:2011xx} (the NNPDF2.3 and ABM11 results
shown are thus the same as in Ref.~\cite{Czakon:2013tha})  .  
Higher twist corrections are shown using the $p_{\rm HT}=1$ curve from Fig.~\ref{fig:ht_corr}, and deuteron corrections are shown
using the MMSTWW curve from Fig.~\ref{fig:deutcorr}.

These cross-sections are shown in Fig.~\ref{fig:lhcxsec}. It is clear that the impact of higher twist and nuclear corrections is negligible, 
both compared to  PDF uncertainties and to the differences between different sets. Based on our previous discussion, it is likely 
that for higher twist corrections this will always be the case, while
nuclear corrections might have a visible impact, up to the half sigma
level on sufficiently exclusive observables which are sensitive to the up-down difference at large $x$ (such as, for example, the charge
asymmetry for very high mass virtual $W^\pm$ production, or, more interestingly, heavy new particles with flavour-dependent
couplings). On the other hand, the use of a FFN scheme has a visible impact, comparable to that of using a smaller dataset which only 
includes DIS data: however, in the latter case the PDF uncertainty automatically increases because of the smaller dataset, while when 
the FFN scheme is adopted an extra theoretical uncertainty should be
added to the result. It is interesting to observe that
Fig.~\ref{fig:lhcxsec} shows that adopting a FFN 
scheme, or fitting only DIS data makes the NNPDF2.3 results closer
to those obtained using the ABM11 set, which is based on a smaller
(mostly DIS) dataset and which uses a FFN scheme. Indeed, we have also
produced a fit using the FFN scheme to DIS data only: the
cross-sections we get, also shown in  Fig.~\ref{fig:lhcxsec}, are in 
suprisingly good agreement with  those obtained using the ABM11 set.

In summary, we have studied the impact of three sources of theoretical uncertainties on PDF determinations: the use 
of a FFN scheme, and the inclusion of higher twist corrections and deuterium nuclear correction. We conclude that, adopting the dataset, methodology, and kinematic cuts of the NNPDF2.3 PDF determination, the impact of the FFN is significant, especially at high scales 
($Q^2\sim M_W^2$),  and that it leads to an extra uncertainty on the results obtained. Higher twist corrections 
have by contrast a negligible impact. Deuterium nuclear corrections have a moderate impact, up to one sigma, but only on the up-down separation 
in the large $x$ region ($x\sim0.3$). Because of the poor theoretical knowledge of these effects, this should again be treated as an extra theoretical
uncertainty. Forthcoming LHC data may help in keeping some of these uncertainties under control, by allowing PDF determinations which 
make no use of data which are subject to nuclear corrections~\cite{Forte:2013wc}.

\vspace{1cm}

{\bf Acknowledgements:} We thank A.~Accardi for providing code for the deuterium corrections of Ref.~\cite{Owens:2012bv}, and
R.~Thorne for providing the deuterium corrections of Ref.~\cite{Martin:2012xx}. J.~R. is supported by a Marie Curie 
Intra--European Fellowship of the European Community's 7th Framework Programme under contract number PIEF-GA-2010-272515. 
S.F. is partially supported by a PRIN 2010 grant. The work of M.U. is supported by the DFG Sonderforschungsbereich/Transregio 9
“Computergest\"utzte Theoretische Teilchenphysik”.


\end{document}